\documentclass[aps,prb,superscriptaddress,showpacs,twocolumn,bibnotes]{revtex4}
\usepackage{graphicx}
\usepackage{amssymb}
\usepackage{amsmath}

\newcommand{\bk}{{\bf k}}
\newcommand{\bq}{{\bf q}}

\newcommand{\bG}{{\bf G}}
\newcommand{\bK}{{\bf K}}
\newcommand{\bM}{{\bf M}}

\newcommand{\bx}{{\bf x}}

\newcommand{\br}{{\bf r}}
\newcommand{\bR}{{\bf R}}

\newcommand{\bdelta}{{\boldsymbol\delta}}

\renewcommand{\Im}{{\mathop{\rm{Im}}\nolimits\,}}
\renewcommand{\Re}{{\mathop{\rm{Re}}\nolimits\,}}

\newcommand{\kB}{k_{\mathrm{B}}}

\newcommand{\nF}{n_{\mathrm{F}}}

\def\bbbone{{\mathchoice {\rm 1\mskip-4mu l} {\rm 1\mskip-4mu l}{\rm 1\mskip-4.5mu l} {\rm 1\mskip-5mu l}}}

\begin{document}

\title{Dynamical polarization of graphene under strain}

\author{F. M. D. Pellegrino}
\affiliation{Dipartimento di Fisica e Astronomia, Universit\`a di Catania,\\
Via S. Sofia, 64, I-95123 Catania, Italy}
\affiliation{CNISM, UdR Catania, I-95123 Catania, Italy}
\affiliation{Scuola Superiore di Catania, Universit\`a di Catania,\\
Via S. Nullo, 5/i, I-95123 Catania, Italy}
\affiliation{CNR-IMM, Z.I. VIII Strada 5, I-95121 Catania, Italy}
\author{G. G. N. Angilella}
\affiliation{Dipartimento di Fisica e Astronomia, Universit\`a di Catania,\\
Via S. Sofia, 64, I-95123 Catania, Italy}
\affiliation{CNISM, UdR Catania, I-95123 Catania, Italy}
\affiliation{Scuola Superiore di Catania, Universit\`a di Catania,\\
Via S. Nullo, 5/i, I-95123 Catania, Italy}
\affiliation{INFN, Sez. Catania, I-95123 Catania, Italy}
\affiliation{CNR-IMM, Z.I. VIII Strada 5, I-95121 Catania, Italy}
\author{R. Pucci}
\affiliation{Dipartimento di Fisica e Astronomia, Universit\`a di Catania,\\
Via S. Sofia, 64, I-95123 Catania, Italy}
\affiliation{CNISM, UdR Catania, I-95123 Catania, Italy}

\date{\today}

\begin{abstract}
We study the dependence of the plasmon dispersion relation of graphene on
applied uniaxial strain. Besides electron correlation at the RPA level, we also
include local field effects specific for the honeycomb lattice. As a consequence
of the two-band character of the electronic band structure, we find two distinct
plasmon branches. We recover the square-root behavior of the low-energy branch,
and find a nonmonotonic dependence of the strain-induced modification of its
stiffness, as a function of the wavevector orientation with respect to applied
strain.
\medskip
\end{abstract} 

\maketitle

\section{Introduction}

Graphene is a two-dimensional single layer of carbon atoms, and can be thought
therefore as the building block of several $sp^2$-bonded carbon allotropes,
ranging from three-dimensional graphite, to one-dimensional nanotubes, to
zero-dimensional fullerenes. Its recent experimental fabrication in the
laboratory \cite{Novoselov:05a} has triggered an enormous outburst of both
experimental and theoretical research. This is justified by the peculiar
electronic and structural properties of graphene
\cite{CastroNeto:08,Abergel:10}, largely due to its reduced dimensionality, as
well as to correlation effects. In particular, its linear quasiparticle
dispersion relation is analogous to that of relativistic massless particles,
obeying Dirac-Weyl equation, thus enabling to study quantum relativistic effects
in a condensed matter system \cite{Zhang:05,Berger:06}.

Most of the unusual electronic properties of graphene are encoded in the
electron polarizability, which has been studied within the Dirac cone
approximation at zero \cite{Gonzalez:99} and finite temperature \cite{Vafek:06}
for pristine graphene, as well as for doped graphene \cite{Wunsch:06,Hwang:07a}.
These results have been recently extended beyond the Dirac cone approximation
\cite{Stauber:10}. The effect of spin-orbit interaction in the
electronic collective excitations of a graphene layer with or without doping has
also been considered in Ref.~\onlinecite{Wang:07}.

Here, we will be concerned on the dynamical polarization of graphene within the
full Brillouin zone of the honeycomb lattice. While electron correlations will
be treated at the RPA level, we will explicitly include local field effects
(LFE) \cite{Adler:62}, which are characteristic of the lattice structure of
graphene. The importance of LFE have been shown to be more important in graphene
than in bulk semiconductors, in connection with the static dielectric properties
of graphene \cite{Schilfgaarde:10}. By discussing the singularities of the
polarizability, we will be able to identify the collective modes of the
correlated electron liquid. We will be mainly interested in the plasmon modes,
which dominate the long wavelength charge density fluctuations. The role of
electron-plasmon interaction in renormalizing the (especially low-energy)
quasiparticle dispersion relation has been emphasized
\cite{Bostwick:07,Brar:10}, and plasmons in graphene are potentially interesting
for applications in nanophotonics \cite{Jablan:09}.

Specifically, we will be interested in the dependence of the plasmon modes on
applied uniaxial strain. This will enable to investigate the interplay between
electronic and structural properties of graphene. It has been even suggested
that nanodevices based on graphene could be engineered on the basis of the
expected strain-induced modifications of the deformed graphene sheet (origami
electronics) \cite{Pereira:09}. Indeed, graphene is also characterized by quite
remarkable mechanical properties. Despite its quasi-two-dimensional character,
it displays an exceptional tensile strength and stiffness \cite{Booth:08}. In
particular, recent \emph{ab initio} calculations
\cite{Liu:07,Cadelano:09,Choi:10,Jiang:10} as well as experiments \cite{Kim:09}
have demonstrated that graphene can sustain elastic deformations as large as
20\%. The possibility of a strain-induced semimetal-to-semiconductor transition,
with the opening of a gap, has been therefore studied
\cite{Gui:08,Pereira:08a,Ribeiro:09,Cocco:10}. It turns out that this critically
depends on the direction of applied strain, as is also confirmed by studies of
the strain effect on the optical conductivity of graphene
\cite{Pellegrino:09b,Pellegrino:09c,optical-lattices}.

The paper is organized as follows. In Sec.~\ref{sec:model} we present our model,
based on a tight-binding description of the graphene electronic band structure.
We will then derive the electronic polarization at RPA level, and explicitly
include local field effects. We will then derive and discuss the various
branches of the plasmon modes along a symmetry contour of the first Brillouin
zone, both numerically and analytically, in the limit of small wavevectors. The
effect of applied uniaxial strain will then be discussed in
Sec.~\ref{sec:strain}. Summary and concluding remarks will be given in
Sec.~\ref{sec:conclusions}.

\section{Model}
\label{sec:model}

\subsection{Tight-binding approximation}

At the tight-binding level of approximation, the Hamiltonian for the graphene
honeycomb lattice can be conveniently written as
\begin{equation}
H = \sum_{\bR,\ell} t_\ell a^\dag (\bR) b(\bR+\bdelta_\ell) + \mathrm{H.c.},
\label{eq:H}
\end{equation}
where $a^\dag (\bR)$ is a creation operator on the position $\bR$ of the A
sublattice, $b(\bR+\bdelta_\ell)$ is a destruction operator on a nearest
neighbor (NN) site $\bR+\bdelta_\ell$, belonging to the B sublattice, and 
$\bdelta_\ell$ are vectors connecting NN sites on different sublattices,
$\bdelta_1 = a(1,\sqrt{3})/2$,  $\bdelta_2 = a(1,-\sqrt{3})/2$,  $\bdelta_3 =
a(-1,0)$, with $a=1.42$~\AA, the equilibrium C--C distance in a graphene sheet
\cite{CastroNeto:08}. In Eq.~(\ref{eq:H}), $t_\ell \equiv t(\bdelta_\ell )$,
$\ell=1,2,3$, is the hopping parameter between two NN sites. In the absence of
strain they reduce to a single constant, $t_\ell \equiv t_0$, with $t_0 =
-2.8$~eV (Ref.~\onlinecite{Reich:02}).

The dispersion relation of the valence ($\lambda=1$) and conduction
bands ($\lambda=2$) are the solutions $E_{\bk\lambda}$ of the
generalized eigenvalue problem
\begin{equation}
H_\bk {\bf u}_{\bk\lambda} = E_{\bk\lambda} S_\bk {\bf u}_{\bk\lambda} ,
\label{eq:eigen}
\end{equation}
where
\begin{subequations}
\begin{eqnarray}
H_\bk &=& \begin{pmatrix} 0 & f_\bk \\ f^\ast_\bk & 0 \end{pmatrix} , \\
S_\bk &=& \begin{pmatrix} 1 & g_\bk \\ g^\ast_\bk & 1 \end{pmatrix} ,
\end{eqnarray}
\end{subequations}
and $\gamma_\bk = \sum_{\ell=1}^3 e^{i\bk\cdot\bdelta_\ell}$, $f_\bk =
\sum_{\ell=1}^3 t_\ell e^{i\bk\cdot\bdelta_\ell}$, $g_\bk = \sum_{\ell=1}^3
s_\ell e^{i\bk\cdot\bdelta_\ell}$ are the usual (complex) structure factor, NN
hopping, and overlap functions in momentum space, respectively. The hopping
parameters $t_\ell$ and overlap parameters $s_\ell$ can be expressed in terms of
appropriate pseudoatomic wave functions, which we here take to be normalized
Gaussian, with standard deviation $\sigma_g$
\cite{Pellegrino:09,Pellegrino:09b}. One finds
\begin{equation}
E_{\bk\lambda} = \frac{-F_\bk \mp \sqrt{F_\bk^2 + 4 G_\bk
|f_\bk|^2}}{2G_\bk} ,
\label{eq:Ek}
\end{equation}
where the minus (plus) sign refers to the valence (conduction) band, and $F_\bk
= g_\bk f_\bk^\ast + g_\bk^\ast f_\bk$ and $G_\bk = 1 - |g_\bk |^2$. In the
following, we shall also use the abbreviation $\xi_{\bk\lambda} = E_{\bk\lambda}
-\mu$, and denote $\bar{\lambda}=2$ for $\lambda=1$, and $\bar{\lambda}=1$ for
$\lambda=2$. Moreover, we also set $\bdelta_A = \bdelta_1 + \bdelta_2 +
\bdelta_3 = {\boldsymbol 0}$, and $\bdelta_B = \bdelta_3$.

A small, albeit nonzero, value of the NN overlap $g_\bk$ has the advantage of
endowing valence and conduction bands with the observed asymmetry. However,
since $g_\bk \approx 0.07 \gamma_\bk \ll 1$ (also under strain, within the range
considered below, in Sec.~\ref{sec:strain}), we can safely retain only linear
corrections to the band dispersions, $E_{\bk\lambda} = \mp |f_\bk | - F_\bk +
\mathcal{O} (g^2_\bk )$, and neglect them altogether in the eigenvectors ${\bf
u}_{\bk\lambda}$.

Our tight-binding approximation is completed by an appropriate choice of the
Bloch wavefunctions. As in Ref.~\onlinecite{Pellegrino:09}, we shall use
$\psi_{\bk\lambda} = N^{-1/2} \sum_j \phi(\br - \bR_j^\lambda )
e^{i\bk\cdot\bR_j^\lambda}$, where $\phi(\br)$ is a Gaussian pseudoatomic
orbital, and $\bR_j^\lambda$ are vectors of the $\lambda=A,B$ sublattices.

We can anticipate, at this stage, that some of the findings of the present study
would not be obtained within the cone approximation. In particular, the
tight-binding approximation allows to include important features of the
electronic band dispersion, such as a finite bandwidth and the occurrence of
Van~Hove singularities. These features will play an essential role in deriving
some of the characteristics of the plasmon dispersion, which is the main goal of
the present work.

\subsection{Local field effects on the electron polarization}

Within linear response theory, plasmon modes can be described as poles of the
density-density correlation function, \emph{i.e.} the polarization. The random
phase approximation (RPA) is then the simplest, infinite order, diagrammatic
procedure to include electron correlations in the dielectric screening giving
rise to the polarization \cite{Giuliani:05}. Besides electron-electron
correlations, another source of $\bk$-space dependence of the dielectric
function is provided by local field effects (LFE) \cite{Schattke:05}. This is
due to the generally atomic consistence of matter and, in the case of solids, to
the periodicity of the crystalline lattice. An account of the LFE on the
dielectric function of crystalline solids dates back at least to the original
paper of Adler \cite{Adler:62} (see also Refs.~\onlinecite{Hanke:74,Hanke:74a}),
and is generalized below to the case of graphene, including both valence and
conduction bands.

We start by considering the polarization, which for a noninteracting system at
finite temperature $T$ reads
\begin{widetext}
\begin{equation}
\Pi^{0} (\bx,\bx^\prime,i\omega_m) = 
\hbar^{-1}(\beta\hbar)^{-1} \sum_{i\omega_n} \sum_{\bk\lambda} \sum_{\bk^\prime\lambda^\prime} 
\psi^\ast_{\bk\lambda} (\bx^\prime) 
{\mathcal G}^0_\lambda (\bk,i\omega_n )
\psi_{\bk\lambda}(\bx) 
\psi^\ast_{\bk^\prime\lambda^\prime} (\bx)
{\mathcal G}^0_{\lambda^\prime} (\bk^\prime , i\omega_n + i\omega_m)
\psi_{\bk^\prime\lambda^\prime}(\bx^\prime) ,
\end{equation}
where ${\mathcal G}^0_\lambda (\bk,i\omega_n ) = (i\omega_n -
\xi_{\bk\lambda}/\hbar)^{-1}$ is the Green's function for the noninteracting system, and
$\hbar\omega_n = (2n+1)\pi\kB T$ [$\hbar\omega_m = 2m\pi\kB T$] denote the
fermionic [bosonic] Matsubara frequencies at temperature $T$, with $\hbar$
Planck's constant and $\kB$ Boltzmann's constant. Fourier transforming into
momentum space, and performing the summation over the Matsubara frequencies, one
finds
\begin{equation}
\Pi^0 (\bq + \bG , -\bq^\prime -\bG^\prime, i\omega_m) =
(2\pi)^2 A_c^{-1} \delta(\bq-\bq^\prime) \frac{1}{N} \sum_{\bk\lambda\lambda^\prime}
T_{\bk\lambda,\bk-\bq\lambda^\prime}(i\omega_m) 
\langle \bk-\bq \lambda^\prime | e^{-i(\bq+\bG)\cdot \hat{\br}} | \bk\lambda
\rangle \langle \bk\lambda | e^{i(\bq+\bG)\cdot\hat{\br}} |
\bk-\bq\lambda^\prime \rangle ,
\label{eq:Pi0}
\end{equation}
\end{widetext}
where
\begin{equation}
T_{\bk\lambda,\bk-\bq\lambda^\prime}(i\omega_m) = \frac{\nF
(\xi_{\bk-\bq\lambda^\prime} ) - \nF (\xi_{\bk\lambda})}{i\hbar\omega_m +
\xi_{\bk-\bq\lambda^\prime} - \xi_{\bk\lambda}} .
\label{eq:T}
\end{equation}
Here, $\nF(\omega)$ is the Fermi function, $A_c = 3\sqrt{3}a^2/2$ is the area of
the Wigner-Seitz cell, $\bq$, $\bq^\prime$ belong to the first Brillouin zone
(1BZ), $\bG$, $\bG^\prime$ are vectors of the reciprocal lattice, and LFE are
embedded in the Adler's weights \cite{Adler:62}
\begin{eqnarray}
\langle \bk-\bq \lambda^\prime | e^{-i(\bq+\bG)\cdot \hat{\br}} | \bk\lambda
\rangle && \nonumber\\
&&\hspace{-4truecm}=\int d^2 \bx e^{-i(\bq+\bG)\cdot \bx}
\psi_{\bk\lambda}(\bx) \psi_{\bk-\bq\lambda^\prime}^\ast (\bx) \nonumber\\
&&\hspace{-4truecm}\simeq 
\frac{1}{2} \left[ (-1)^{\lambda-\lambda^\prime} +
e^{i(\theta_{\bk-\bq} - \theta_\bk ) -i\bG\cdot\bdelta_3} \right]
e^{-\sigma_g^2 | \bq + \bG |^2 /4 } ,
\end{eqnarray}
where in the last line only the onsite overlap between pairs of pseudoatomic
orbitals, centered on either sublattices, has been retained, on account of their
localized character, we have retained only the lowest (zeroth) order
contributions in the overlap function $g_\bk$, and $e^{i\theta_\bk} = - f_\bk /
|f_\bk |$. Using a more compact notation, one may also write
\begin{eqnarray}
\Pi^0 (\bq + \bG , -\bq^\prime -\bG^\prime, i\omega_m) &=& 
(2\pi)^2 A_c^{-1} \delta(\bq-\bq^\prime) \nonumber\\
&&\hspace{-2.5truecm}\times
\sum_{\alpha\beta} \rho_{\bq\alpha} (\bG) Q^0_{\alpha\beta} (\bq,i\omega_m)
\rho^\ast_{\bq\beta} (\bG^\prime) ,
\end{eqnarray}
where 
\begin{eqnarray}
Q^0_{\alpha\beta} (\bq,i\omega_m) &=& \frac{1}{N} \sum_{\bk\lambda\lambda^\prime}
u_{\bk\lambda}^\alpha u_{\bk\lambda}^{\beta\ast}
u_{\bk-\bq\lambda^\prime}^{\alpha\ast} u_{\bk-\bq\lambda^\prime}^\beta
\nonumber\\
&&\times
T_{\bk\lambda,\bk-\bq\lambda^\prime}(i\omega_m) ,
\label{eq:Q0}
\end{eqnarray}
with $u_{\bk\lambda}^\alpha$ the components of ${\bf u}_{\bk\lambda}$
($\alpha=1,2$), and
\begin{equation}
\rho_{\bq\alpha} (\bG) = \exp(-i\bG\cdot\bdelta_\alpha - \sigma_g^2 |\bq+\bG|^2
/4)
\end{equation}
are the LFE weights. The continuum limit is recovered when $\bG=\bG^\prime=0$.

Many-body correlations are then included within RPA, yielding a renormalized
polarization
\begin{eqnarray}
\Pi (\bq + \bG , -\bq^\prime -\bG^\prime, i\omega_m) &=& 
(2\pi)^2 A_c^{-1} \delta(\bq-\bq^\prime) \nonumber\\
&&\hspace{-2.5truecm}\times
\sum_{\alpha\beta} \rho_{\bq\alpha} (\bG) Q_{\alpha\beta} (\bq,i\omega_m)
\rho^\ast_{\bq\beta} (\bG^\prime) ,
\label{eq:Pi}
\end{eqnarray}
where now
\begin{equation}
Q (\bq,i\omega_m) = g_s Q^0 (\bq,i\omega_m) [ \bbbone - g_s A_c^{-1} V(\bq) Q^0 (\bq,i\omega_m)]^{-1} ,
\label{eq:Q}
\end{equation}
where matrix products are being understood and $g_s =2$ is a factor for spin
degeneracy, and
\begin{equation}
V_{\alpha\beta} (\bq) = \sum_{\bG^{\prime\prime}} \rho_{\bq\alpha}^\ast
(\bG^{\prime\prime}) V_0 (\bq+\bG^{\prime\prime}) \rho_{\bq\beta}
(\bG^{\prime\prime} )
\label{eq:Coulomb}
\end{equation}
is the renormalized Coulomb potential, $V_0(\bq) = e^2/(2\varepsilon_0
\varepsilon_r q)$, now a matrix over band indices.  Here, $\varepsilon_r =
(\varepsilon_{r1} + \varepsilon_{r2})/2$ denotes te average relative dielectric
constants of the two media surrounding the graphene layer. These are air for
suspended graphene ($\varepsilon_{r1} = \varepsilon_{r2} = \varepsilon_r = 1$).
In the case of a stronger dielectric ssubstrate, we expect therefore a softening
of the correlation effects on the plasmon frequency. It is relevant to note
that the renormalized potential already includes LFE. 

\subsection{Plasmons}
\label{ssec:plasmons}

Plasmons are defined as collective excitations of the electron liquid
corresponding to poles of the retarded polarization,
\begin{equation}
\Pi(\bq,\omega) \equiv \Pi(\bq,-\bq,i\omega_m \to \omega+i0^+),
\label{eq:PiR}
\end{equation}
where $\bq\in\mbox{1BZ}$. Here and in
what follows we shall restrict to the case $\bG = \bG^\prime =0$. Indeed, it is
apparent from the definition of $\Pi(\bq,\omega)$ that its poles can only arise
from the vanishing of $\det [ \bbbone - V(\bq) Q^0 (\bq,\omega)]$ in
Eq.~(\ref{eq:Q}), which already contains LFE via the renormalized Coulomb
potential, Eq.~(\ref{eq:Coulomb}). We therefore define the dispersion relation
$\omega_\ell (\bq)$ of the $\ell$-th plasmon branch as
\begin{equation}
\Pi^{-1} \left( \bq, \omega_\ell (\bq) \right) = 0.
\label{eq:omega}
\end{equation}
This clearly involves vanishing of both real and imaginary parts of the inverse
polarization. It will be useful to define the dispersion relation
$\tilde{\omega}_\ell (\bq)$ of damped plasmons through
\begin{equation}
\Re \left[ \Pi^{-1} \left( \bq, \tilde{\omega}_\ell (\bq) \right) \right]= 0.
\label{eq:dampedomega}
\end{equation}
Correspondingly, the inverse lifetime $\tau^{-1} (\bq,\omega)$ of such damped
plasmons is proportional to $-\Im \Pi(\bq,\omega)$, for
$\omega=\tilde{\omega}_\ell (\bq)$.

\begin{figure}[t]
\centering
\includegraphics[width=\columnwidth]{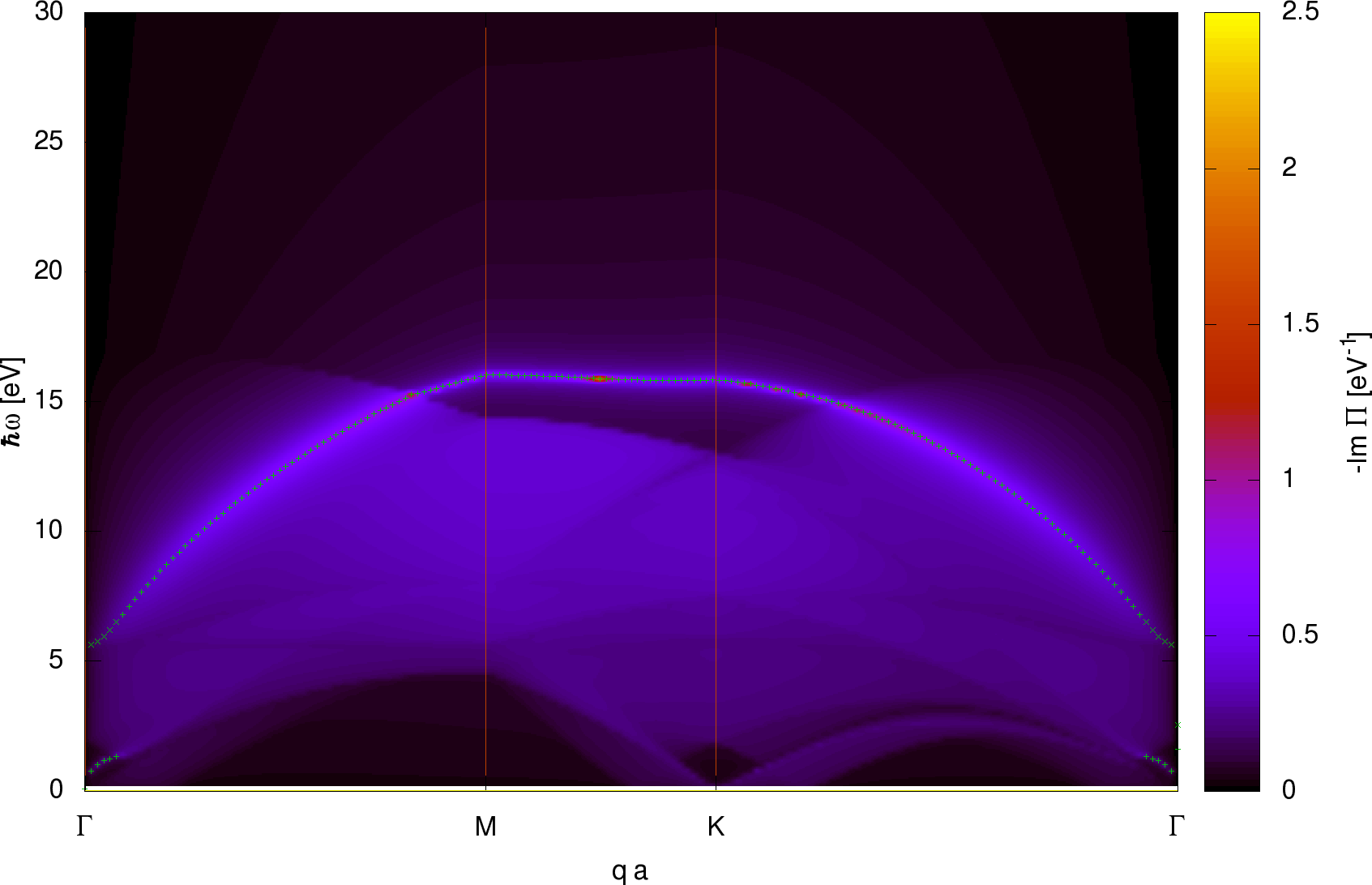}
\includegraphics[width=\columnwidth]{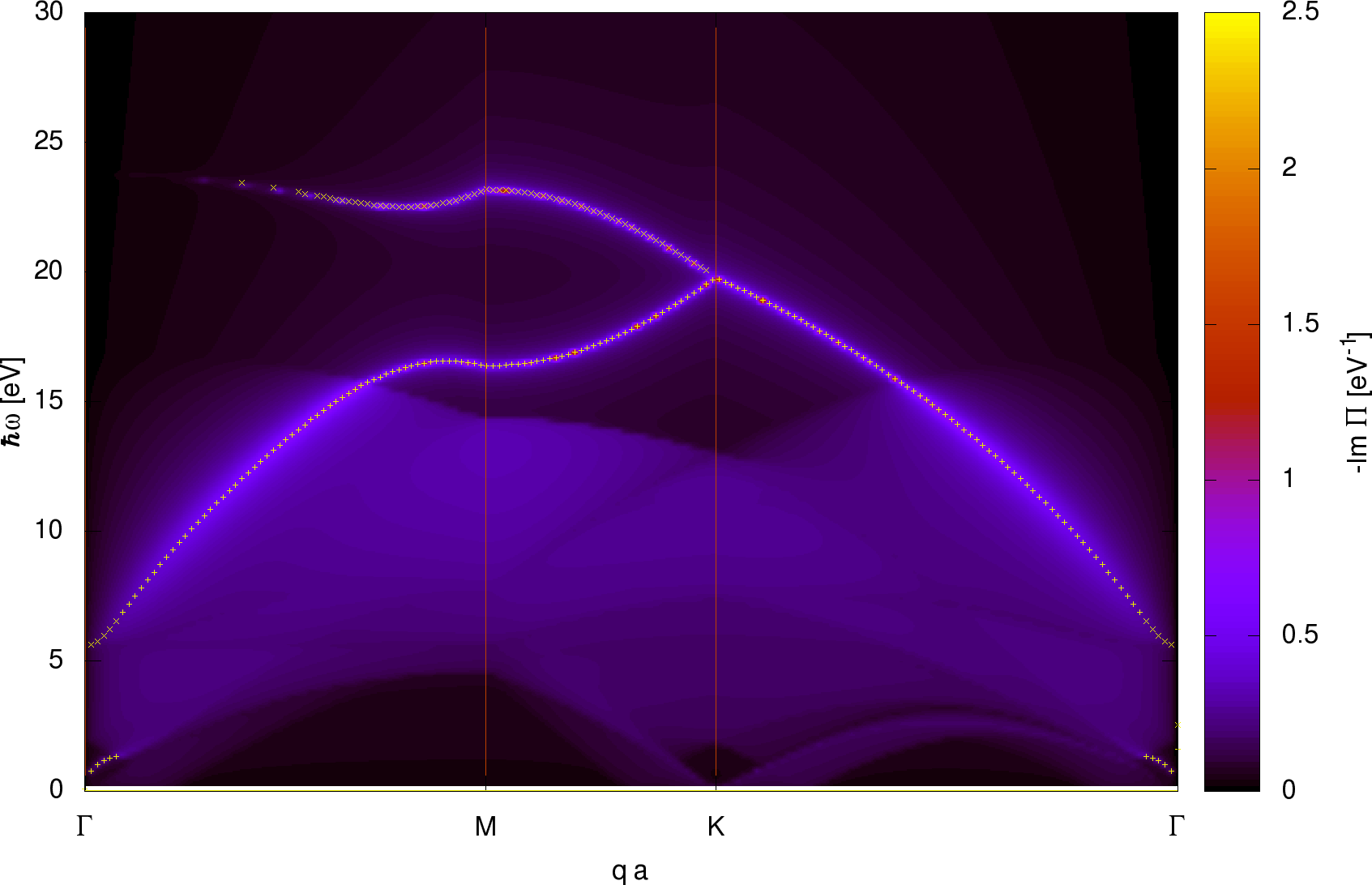}
\caption{(Color online) Plasmon dispersion relation for doped graphene
($\mu=1$~eV) at finite temperature ($T=3$~K), not including (top panel) and
including (bottom panel) LFE. Results are shown along a symmetry contour in the
1BZ, with $\Gamma=(0,0)$, $\bM=(2\pi/3a,0)$, and
$\bK=(2\pi/3a,2\pi/3\sqrt{3}a)$. Frequencies $\omega$ are in eV. The shaded
background is a contour plot of $-\Im\Pi(\bq,\omega)$ (arbitrary scale), while
continuous lines are the dispersion relation of damped plasmons,
$\tilde{\omega}_\ell (\bq)$, Eq.~(\ref{eq:dampedomega}), is shown as a dotted
line.}
\label{fig:lfe}
\end{figure}

Fig.~\ref{fig:lfe} shows our numerical results for the plasmon dispersion
relation in doped graphene ($\mu=1$~eV) at finite temperature ($T=3$~K) along
a symmetry contour in the 1BZ, without LFE [$\bG^{\prime\prime}=0$ in
Eq.~(\ref{eq:Coulomb}), top panel] and including LFE (bottom panel). At small
wavevectors and low frequencies, one recognizes a square-root plasmon mode
$\omega_1 (\bq)\sim \sqrt{q}$, typical of a 2D system \cite{Giuliani:05}. This
is in agreement with earlier studies of the dynamical screening effects in
graphene at RPA level, employing an approximate conic dispersion relation for
electrons around the Dirac points \cite{Wunsch:06,Hwang:07a}. Such a result has
been confirmed also for a tight-binding band \cite{Hill:09,Stauber:10}, and is
here generalized with the inclusion of LFE. The effect of spin-orbit
interaction can be neglected, in the case of sufficiently large chemical
potential \cite{Wang:07}, as is here the case.

The high energy ($5-20$~eV) pseudo-plasmon mode, extending throughout the whole
1BZ, is rather associated with a logarithmic singularity of the bare
polarization $Q^0 (\bq,\omega)$ in Eq.~(\ref{eq:Q}), and therefore does not
correspond to a true pole of the polarization. This collective mode can be
related to an interband transition between the Van~Hove singularities in the
valence and conduction bands of graphene, and has been identified with a
$\pi\to\pi^\ast$ transition \cite{Gass:08,Stauber:10}.

At large wavevectors, specifically along the zone boundary between the $\bM$ and
the $\bK$ (Dirac) points, full inclusion of LFE determines the appearance of a
second, high-frequency ($20-25$~eV), optical-like plasmon mode $\omega_2 (\bq)$,
weakly dispersing as $q\to 0$. Multiple plasmon modes are a generic consequence
of the possibility of interband transitions, whenever several such bands are
available. This is \emph{e.g.} the case of quasi-2D quantum wells (2DQW), whose
energy spectrum is characterized by quantized levels in the direction
perpendicular to the plane of the well, while electrons can roam freely within
the plane \cite{Giuliani:05}. In this case, collective modes arise as zeroes of
the determinant of the dielectric function. At low temperatures, at most the two
lowest subbands need to be considered. One usually obtains a longitudinal
`acoustic' mode associated to intrasubband coupling, and a transverse `optical'
mode associated to intersubband coupling \cite{Ullrich:02}. Such a situation is
here paralleled by the case of graphene, the role of the two subbands of 2DQW
being here played by the valence and conduction bands, touching at the Dirac
points in the neutral material. It should be noticed that the plasmon mode due
to interband coupling is exactly suppressed when LFE are neglected. In 2DQW, the
discrete nature of the electronic subbands is due to the real-space confinement
of the electron liquid in the direction perpendicular to the plane, \emph{i.e.}
to the {\em quasi}-2D character of the quantum well. In graphene, the origin of
the two bands ultimately lies in the specific lattice structure of this
material. Therefore, the high-energy, `optical' plasmon mode disappears in the
absence of LFE (Fig.~\ref{fig:lfe}, top panel), as expected whenever the lattice
structure of graphene is neglected. 
In other words, while in the absence of LFE only scattering processes with
momenta within the 1BZ are considered, LFE allow to include all scattering
processes with arbitrarily low wavelengths, thereby taking into account the
discrete nature of the crystalline lattice. Such a structure needs not be
considered in the case of a 2DQW.
Our finding of a high-energy `optical'
plasmon branch, as generic consequence of the two-band electronic structure of
graphene, should stimulate further investigation of the electronic collective
modes in graphene \cite{Hill:09,Eberlein:08}, in view of the role of
electron-electron correlations in interpreting electron spectroscopy for
interband transitions \cite{Polini:08}.

\subsection{Asymptotic behaviors}
\label{sec:asymptotics}

In certain limiting regimes, one may derive the asymptotic behavior of the
polarization in close form. At low energies ($\hbar\omega\lesssim |t|$) and
small wavevectors ($q\to0$, \emph{i.e.} $q\sigma_g \ll 1$), LFE can be
neglected. The matrix product entering the definition of
the polarization through Eq.~(\ref{eq:Q}) then reduces to
\begin{eqnarray}
g_s A_c^{-1} V(\bq) Q^0 (\bq,\omega) &=& 
g_s A_c^{-1} V_0 (\bq) \sum_{\alpha\beta} Q^0_{\alpha\beta} (\bq,\omega) 
\nonumber\\
&=& 
\frac{\tilde{V}_0}{qa} 
\frac{1}{N} \sum_{\bk\lambda} \delta_T (\xi_{\bk\lambda} )
\left(\frac{\bq\cdot\nabla_\bk E_{\bk\lambda}}{\hbar\omega} \right)^2 ,
\label{eq:VQa}
\end{eqnarray}
where $\tilde{V}_0 = g_s (8\pi/3\sqrt{3} ) (a_0/a)$~Ry, $a_0$ being Bohr's
radius, and  $\delta_T (\epsilon) \equiv -\partial \nF
(\epsilon)/\partial\epsilon \to \delta(\epsilon)$, as $T\to0$. In the latter
limit, the $\delta$-function effectively restricts the integration over
wavevectors along the Fermi line. Whenever the cone approximation holds
(\emph{i.e.,} for sufficiently low chemical potential and strain; see
Sec.~\ref{sec:strain}), this can be taken as the constant-energy ellipse in
Eq.~(17) of Ref.~\onlinecite{Pellegrino:09b}. The $\bk$-integration in
Eq.~(\ref{eq:VQa}) can then be performed analytically, and the retarded
polarization, Eq.~(\ref{eq:PiR}), then reads
\begin{equation}
\Pi(\bq,\omega) \approx \frac{g_s A_c^{-1} \tilde{V}_0^{-1} \tilde{\omega}_1^2 q^2
a^2}{\hbar^2{\omega^+}^2 - \hbar^2 \omega_1^2 (\bq)} ,
\label{eq:Pia}
\end{equation}
where $\omega^+ \equiv \omega+i0^+$, and
\begin{equation}
\hbar\tilde{\omega}_1 = \left( \frac{1}{2} \tilde{V}_0 \rho(\mu) \right)^{1/2} 
|\nabla_\bq E_{\bq2} /a| ,
\label{eq:alpha1}
\end{equation}
with $g_s\rho(\mu)$ the density of states (DOS) at the Fermi level. To leading
order in $qa$, from Eq.~(\ref{eq:Pia}) one thus obtains
\begin{equation}
\omega_1 (\bq) \approx \tilde{\omega}_1 \sqrt{qa} 
\label{eq:omegaasymp}
\end{equation}
for the acoustic-like plasmon dispersion relation. One thus recovers the
square-root behavior of the plasmon dispersion relation, as is typical in 2D
electron systems \cite{Giuliani:05}. Moreover, one recovers the dependence of
the coefficient $\tilde{\omega}_1 \sim n^{1/4}$ on the carrier density $n$,
rather than $\sim n^{1/2}$, as is the case for a parabolic dispersion relation
of the quasiparticles \cite{Hwang:07a,DasSarma:09}. 
The acoustic-like plasmon mode may be related to the Drude weight
\cite{Polini:09}, thus enabling the observation of strain effects from optical
measurements \cite{Pellegrino:10c}.
In the case of graphene on a dielectric substrate ($\varepsilon_r > 1$), one has
a reduction of $\tilde{\omega}_1$, thus a softening of the plasmon mode.
From Eq.~(\ref{eq:Pia}) one
may also read off the imaginary part of the retarded polarization, which close
to the `acoustic' plasmon mode [$\omega\sim\omega_1(\bq)$] reads
\begin{equation}
\Im \Pi(\bq,\omega^+ ) \approx -\frac{\pi}{2} g_s A_c^{-1} \tilde{V}_0^{-1/2}
\tilde{\omega}_1 (qa)^{3/2} \delta\left(\omega-\omega_1(\bq)\right).
\label{eq:ImPia1}
\end{equation}

We now turn to the asymptotic behavior of the second branch of the plasmonic
spectrum, $\omega_2 (\bq)$. We have already established that it displays an
optical-like character, with $\omega_2 (\bq)\to\omega_2 (0)$, as $q\to0$.  Here,
$\omega_2 (0)$ is greater than the distance between the top of the conduction
band and the bottom of the valence band. At small wavevectors, it is useful to
consider the expansions of the relevant terms in Eq.~(\ref{eq:Q}), which to
leading order in $q_i$ ($i=x,y$) read
\begin{subequations}
\begin{eqnarray}
Q^0_{AA} (\bq,\omega) &\approx& Q_{AA} (0,\omega) + \sum_{ij} q_i y_{ij}(\omega)
q_j ,\\
Q^0_{AB}(\bq,\omega) &\approx& -Q_{AA} (0,\omega) + \sum_{ij} q_i z_{ij}(\omega)
q_j ,
\end{eqnarray}
\end{subequations}
where $y_{ij} (\omega)$, $z_{ij} (\omega)$ are real valued functions of the
frequency $\omega$, and
\begin{equation}
Q^0_{AA} \left(0,\omega_2(0)\right) = \frac{1}{4N} \sum_{\bk\lambda}
\frac{\nF(\xi_{\bk\bar{\lambda}}) - \nF(\xi_{\bk\lambda})}{\omega_2(0) +
\xi_{\bk\bar{\lambda}} - \xi_{\bk\lambda}} .
\label{eq:Q0AA}
\end{equation}
The asymptotically constant value of the optical-like plasmon frequency is then
implicitly given by
\begin{equation}
1-4Q^0_{AA} \left(0,\omega_2(0)\right) g_s A_c^{-1} 
\sum_\bG V(\bG)\sin^2 \left(\frac{1}{2} \bG\cdot\bdelta_3 \right) =0,
\label{eq:Q0AA1}
\end{equation}
whereas the imaginary part of the retarded polarization, close to the second
plasmon branch [$\omega\sim\omega_2(0)$], to leading order in $q$, reads
\begin{eqnarray}
\Im \Pi (\bq,\omega^+ ) &\approx& -\pi g_s A_c^{-1}
\left|\frac{1}{4N} \sum_{\bk\lambda}
\frac{\nF(\xi_{\bk\bar{\lambda}}) - \nF(\xi_{\bk\lambda})}{(\omega_2(0) +
\xi_{\bk\bar{\lambda}} - \xi_{\bk\lambda})^2}\right|^{-1}
\nonumber\\
&&\hspace{-1.5truecm}\times
\sum_{ijhk} q_i q_h (z_{ij} - y_{ij}) 
(z_{hk} + y_{hk}) q_j q_k 
\delta\left(\omega - \omega_2 (0) \right) .
\label{eq:ImPia2}
\end{eqnarray}
In particular, it follows that the spectral weight of $\Im\Pi$ close to
$\omega_2 (0)$ decreases as $\sim q^4$, as $q\to0$, rather than as $\sim
q^{3/2}$, as is the case for the acoustic-like plasmon mode,
Eq.~(\ref{eq:ImPia1}). This justifies the reduced spectral weight associated
with the second plasmon branch at small wavevector in Fig.~\ref{fig:lfe}.
In the case of graphene on a dielectric substrate ($\varepsilon_r >1$),
inspection of Eqs.~(\ref{eq:Q0AA}) and (\ref{eq:Q0AA1}) yields a reduction of
$\omega_2(0)$.

\section{Effect of strain on the plasmon dispersion relation}
\label{sec:strain}

We now turn to consider the effect of strain on the plasmon dispersion relation.
As in Refs.~\onlinecite{Pereira:08a,Pellegrino:09b}, applied uniaxial strain can
be modeled by explicitly considering the dependence on the strain tensor
$\boldsymbol{\varepsilon}$ of the tight-binding parameters $t_\ell =
t(\bdelta_\ell)$ through the vectors $\bdelta_\ell$ connecting two NN sites
($\ell=1,2,3$). A linear dependence of $\bdelta_\ell$ on
$\boldsymbol{\varepsilon}$ is justified in the elastic limit. Such an assumption
is however quite robust, due to the extreme rigidity of graphene
\cite{Booth:08}, and is supported by \emph{ab initio} calculations
\cite{Cadelano:09,Jiang:10}.

Below, the strain tensor $\boldsymbol{\varepsilon}$ will be parametrized by a
strain modulus $\varepsilon$, and by the angle $\theta$ between the direction of
applied strain and the $x$ axis in the lattice coordinate system. Specifically,
one has $\theta=0$ [\emph{resp.,} $\theta=\pi/6$] for strain applied along the
armchair [\emph{resp.,} zig-zag] direction.

The main effect of applied strain on the band dispersion relation is that of
shifting the location of the Dirac points $\pm\bk_D$ from their position
$\pm\bK$ at $\varepsilon=0$. While valence and conduction bands vanish linearly
as $\bq\to\pm\bk_D$ for moderately low applied strain, such an approximation
breaks down at a critical value of the strain modulus $\varepsilon$, depending
on the direction $\theta$ of applied strain, when $\pm\bk_D$ tends to either
midpoint $\bM_\ell$ of the 1BZ border. This has been described in terms of an
electronic topological transition (ETT), since it is accompanied by a change of
topology of the Fermi line \cite{Pellegrino:09b}.

\begin{figure}[t]
\begin{center}
\includegraphics[width=0.45\columnwidth]{angilfe.png}
\includegraphics[width=0.45\columnwidth]{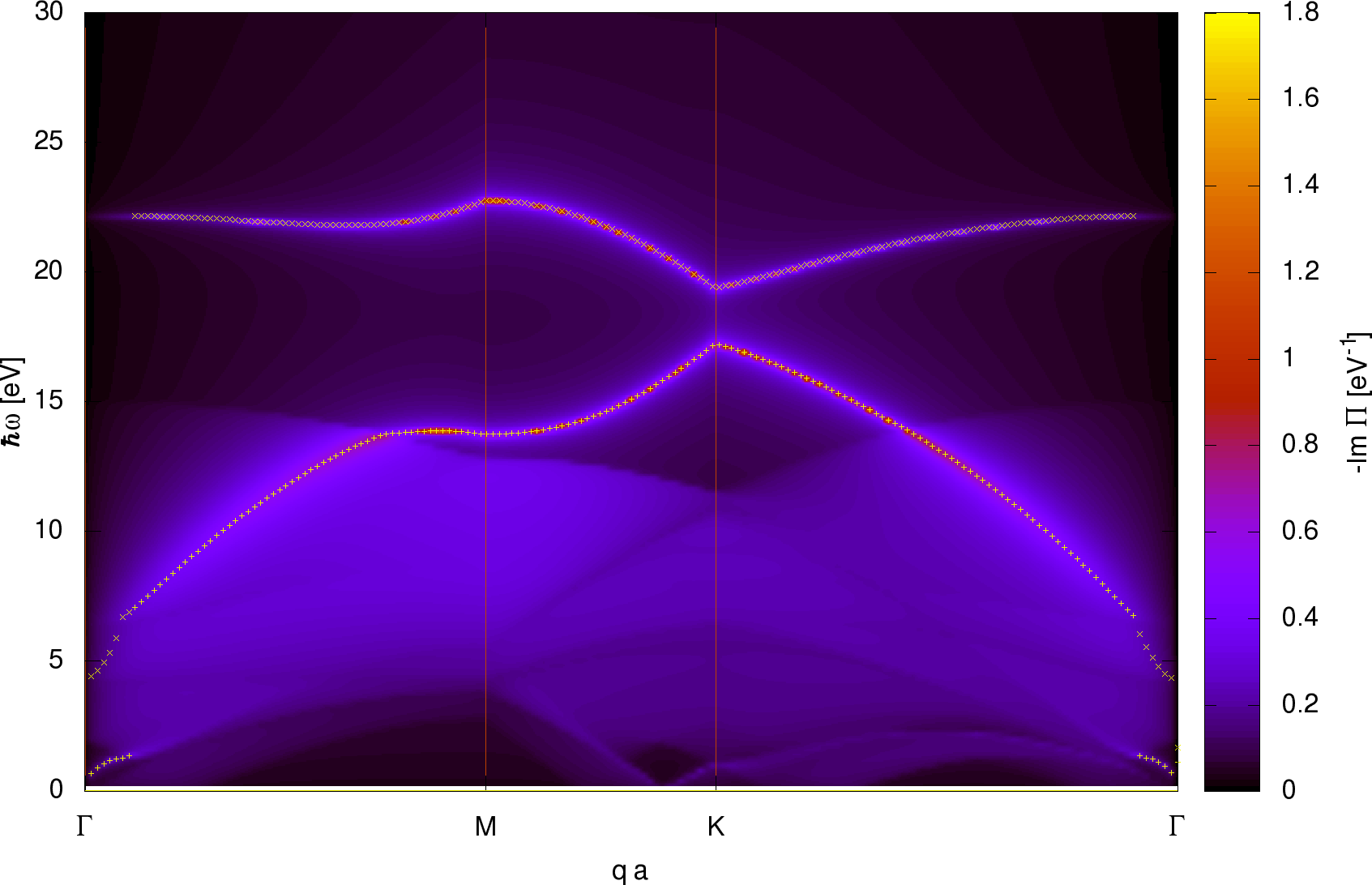}\\
\includegraphics[width=0.45\columnwidth]{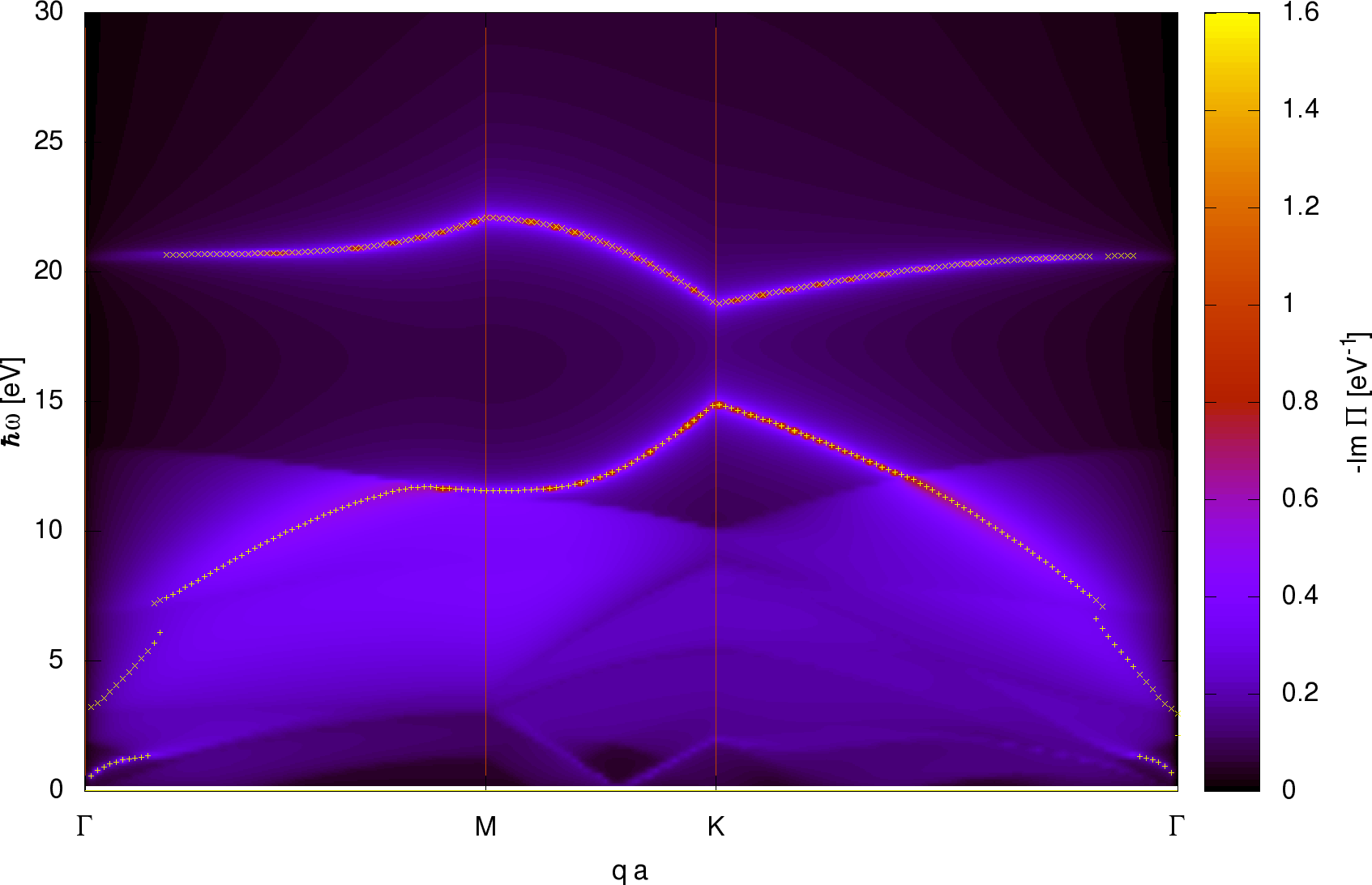}
\includegraphics[width=0.45\columnwidth]{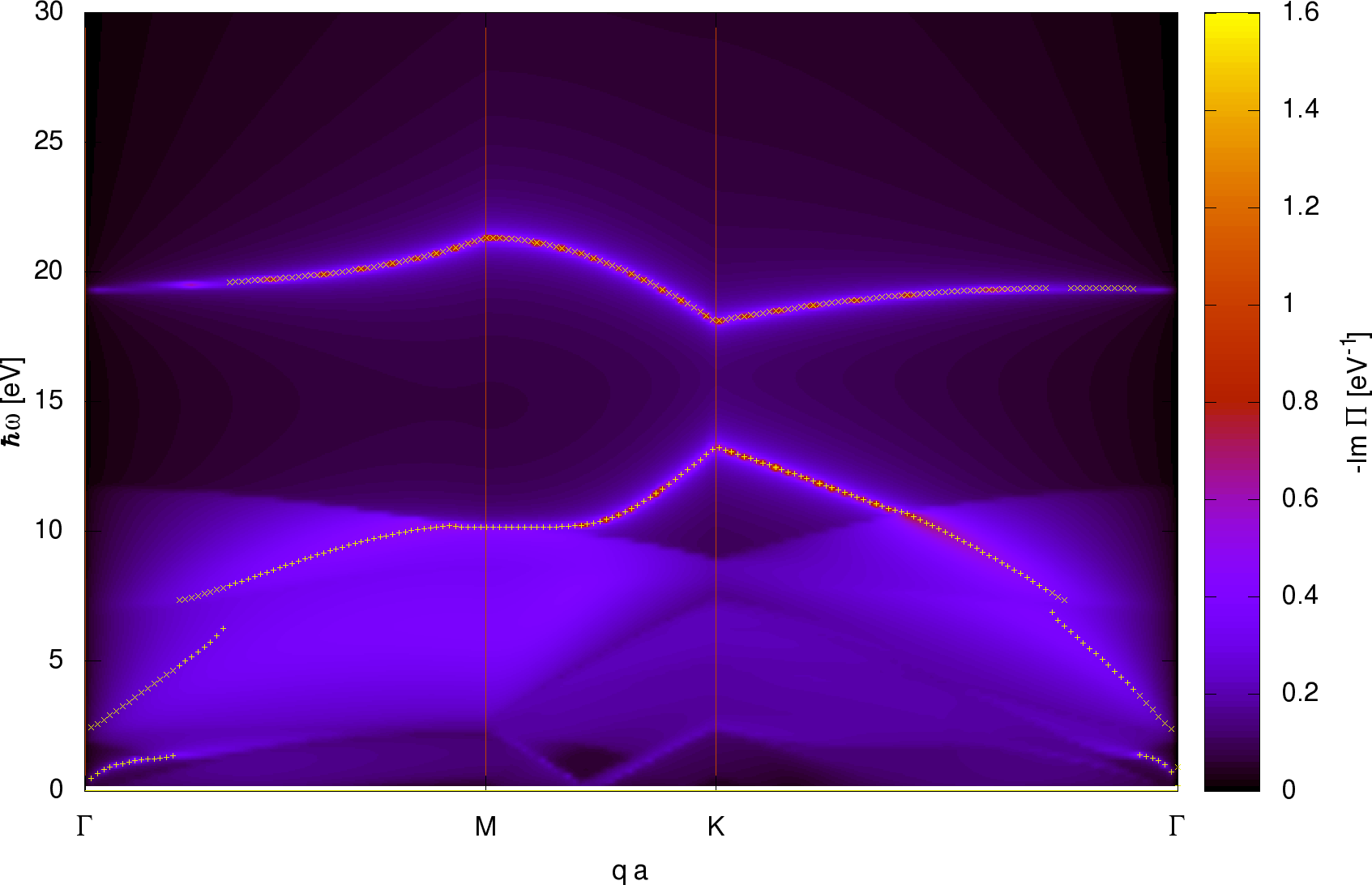}
\end{center}
\caption{(Color online) Plasmon dispersion relation for doped graphene at finite
temperature ($\mu=1$~eV, $T=3$~K), including LFE, with strain applied along
the $\theta=0$ (armchair) direction. Strain increases (from left to right, from
top to bottom) as $\varepsilon = 0$, $0.075$, $0.175$, $0.275$.}
\label{fig:strain0}
\end{figure}

Fig.~\ref{fig:strain0} shows the dispersion relation of the plasmon branches
studied in Sec.~\ref{ssec:plasmons}, including LFE, along a symmetry contour of
the 1BZ, for strain applied along the armchair direction ($\theta=0$), with
increasing strain modulus ($\varepsilon=0-0.275$). The low-frequency, `acoustic'
plasmon mode $\omega_1 (\bq)$ is not qualitatively affected by the applied
strain. In particular, the dominant square-root behavior is independent with
respect to the opening of a gap. On the other hand, one observes an increase of
spectral weight associated with the high-frequency, `optical' plasmon mode
$\omega_2 (\bq)$ at small wavevectors. The overall flattening of the second
plasmon branch over the symmetry contour under consideration can be traced back
to the strain-induced shrinking of both valence and conduction bands. We also
note the formation of a gap between $\omega_2 (\bq)$ and the pseudo-plasmon mode
corresponding to a logarithmic singularity in $Q^0 (\bq,\omega)$ at $\bq=\bK$.

\begin{figure}[t]
\begin{center}
\includegraphics[width=0.45\columnwidth]{angilfe.png}
\includegraphics[width=0.45\columnwidth]{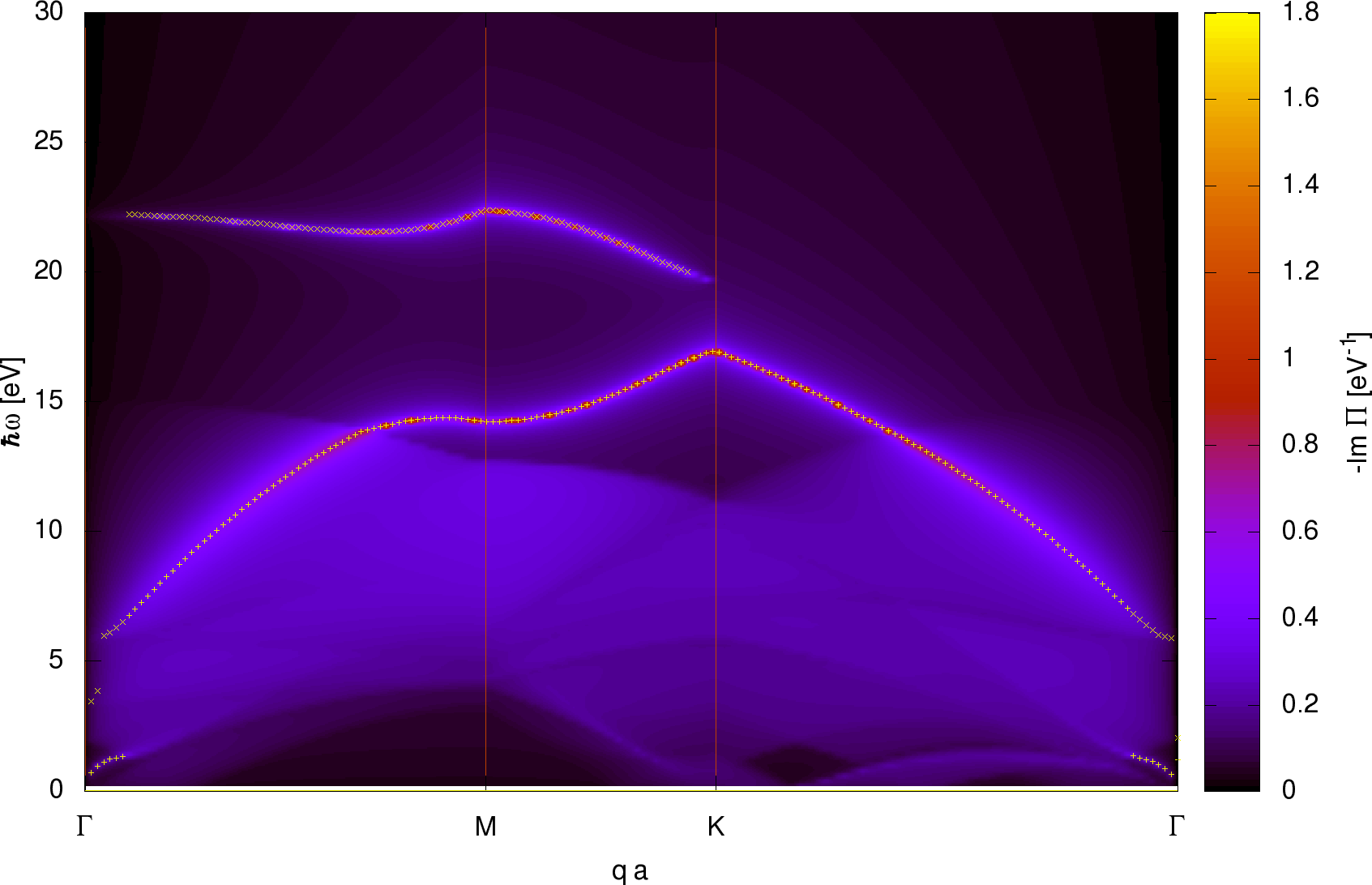}\\
\includegraphics[width=0.45\columnwidth]{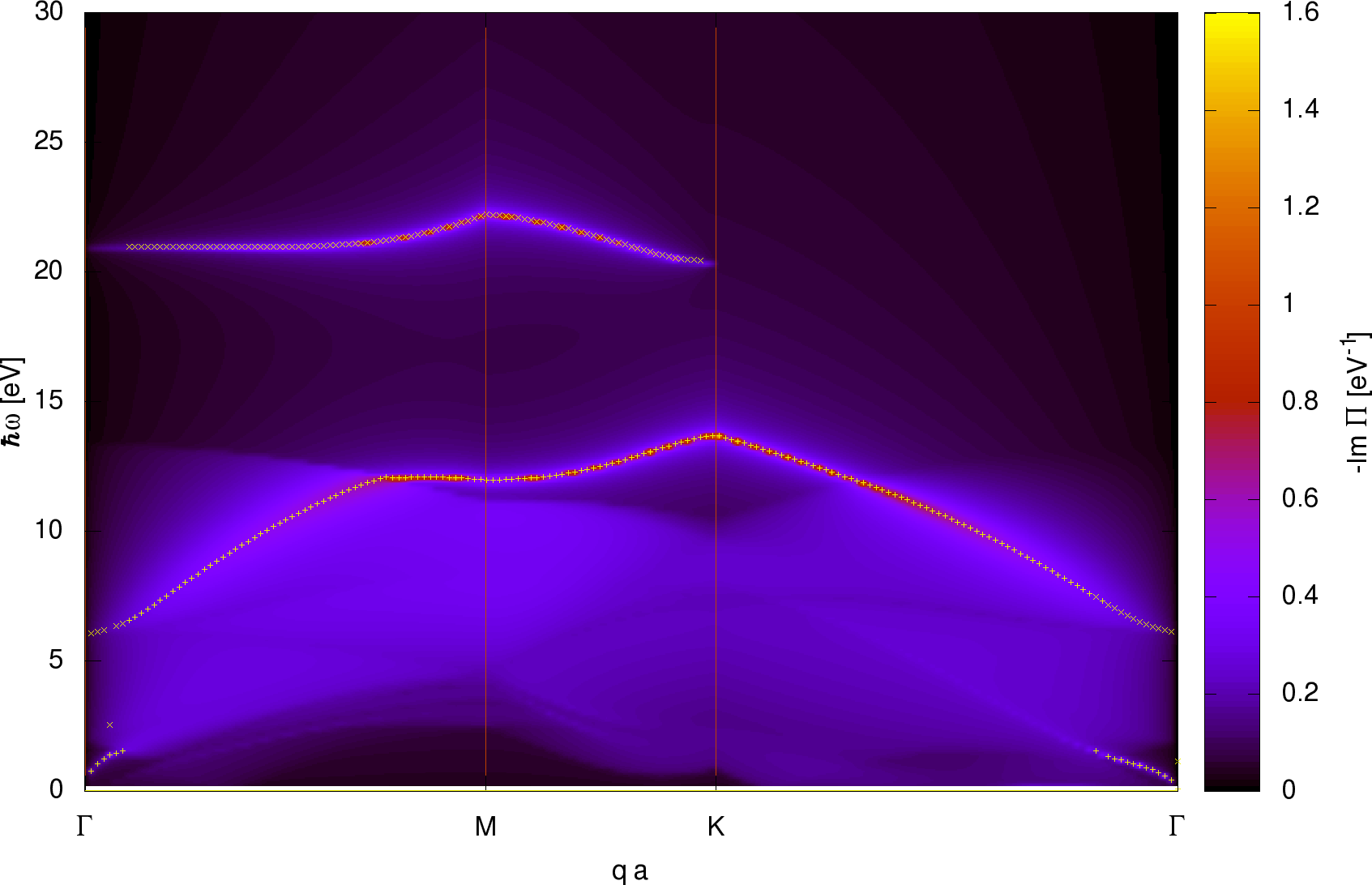}
\includegraphics[width=0.45\columnwidth]{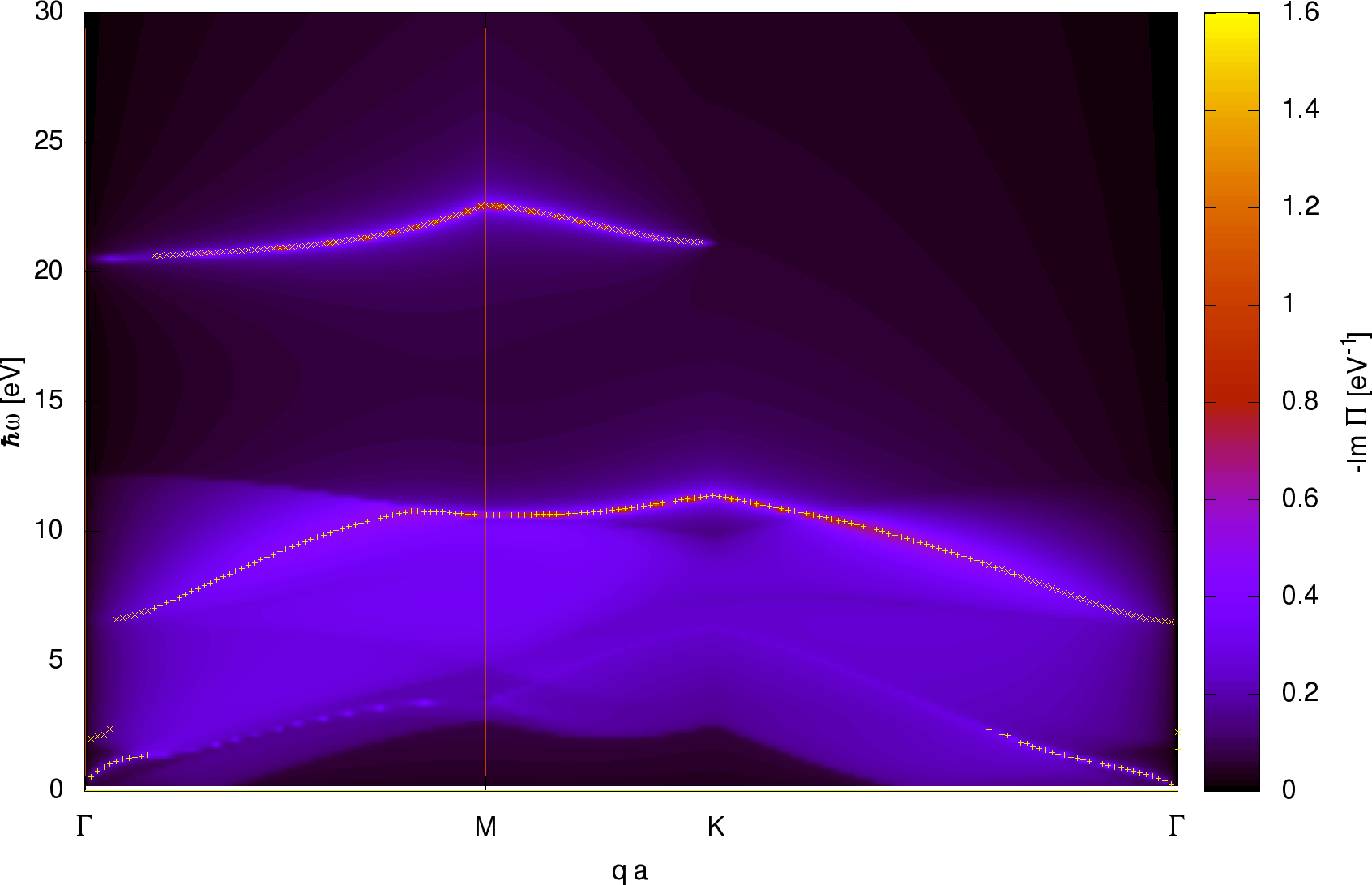}
\end{center}
\caption{(Color online) Plasmon dispersion relation for doped graphene at finite
temperature ($\mu=1$~eV, $T=3$~K), including LFE, with strain applied along
the $\theta=\pi/6$ (zig-zag) direction. Strain increases (from left to right,
from top to bottom) as $\varepsilon = 0$, $0.075$, $0.175$, $0.275$.}
\label{fig:strain30}
\end{figure}

\begin{figure}[t]
\begin{center}
\includegraphics[width=0.45\columnwidth]{angilfe.png}
\includegraphics[width=0.45\columnwidth]{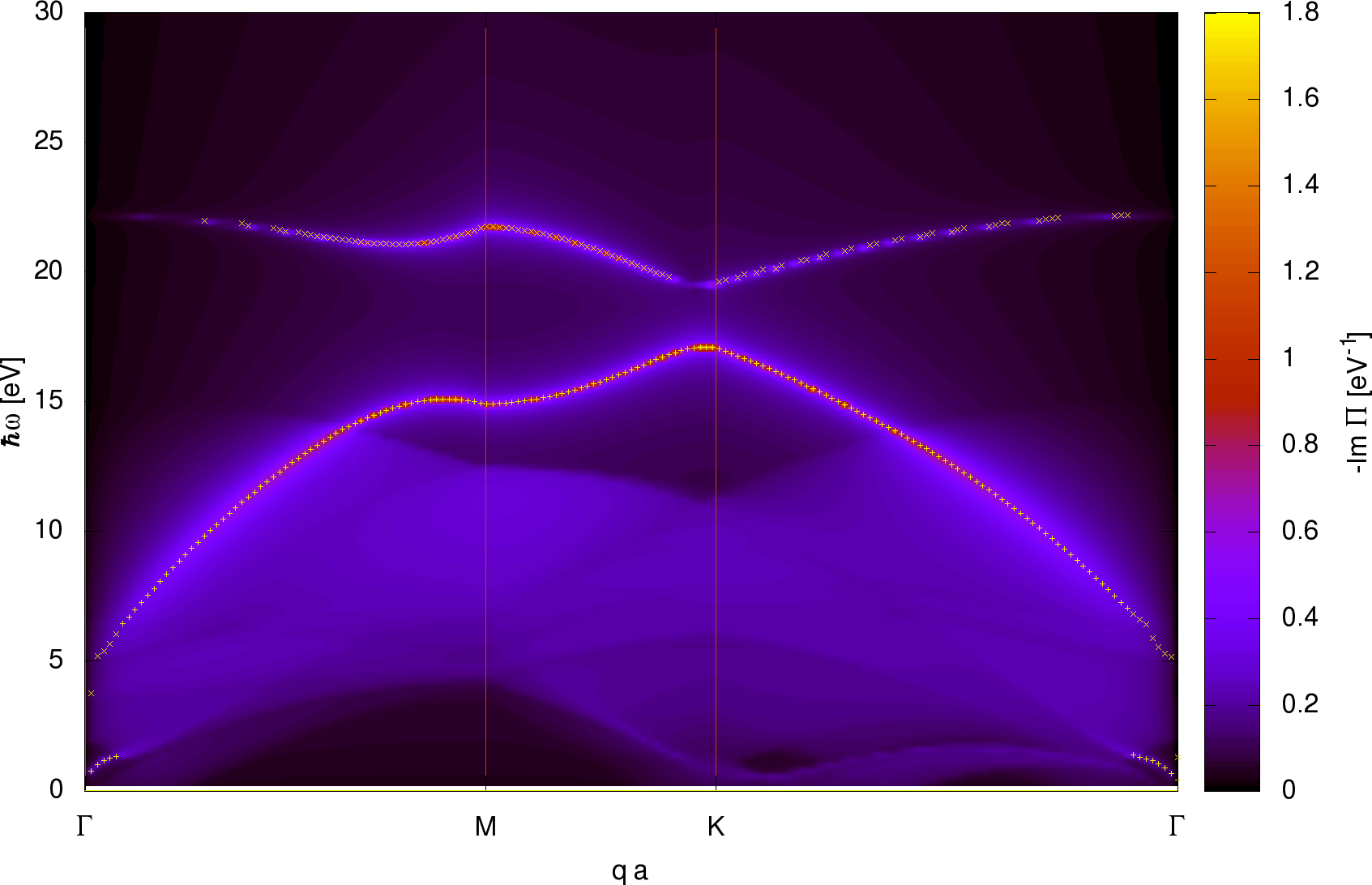}\\
\includegraphics[width=0.45\columnwidth]{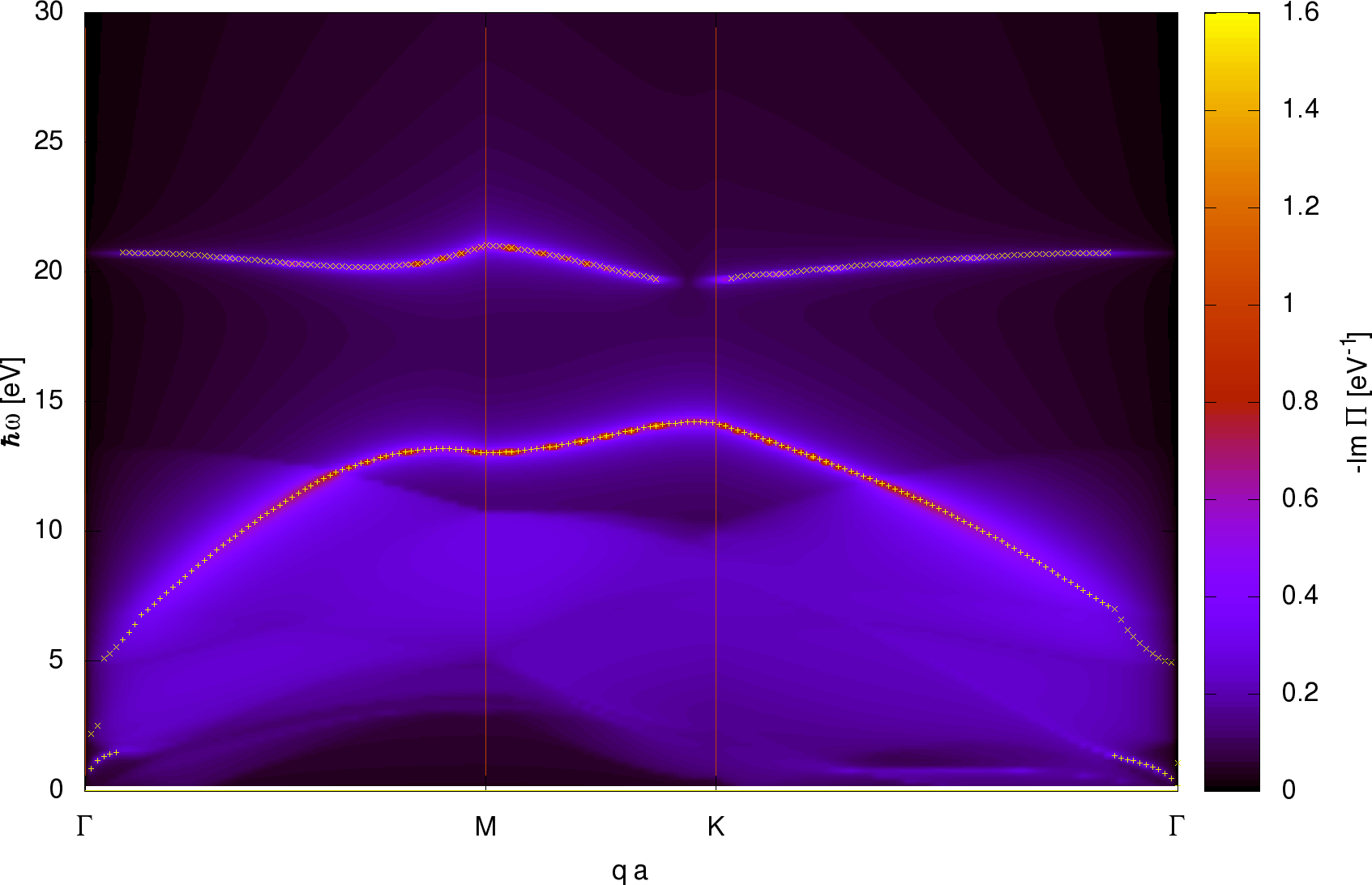}
\includegraphics[width=0.45\columnwidth]{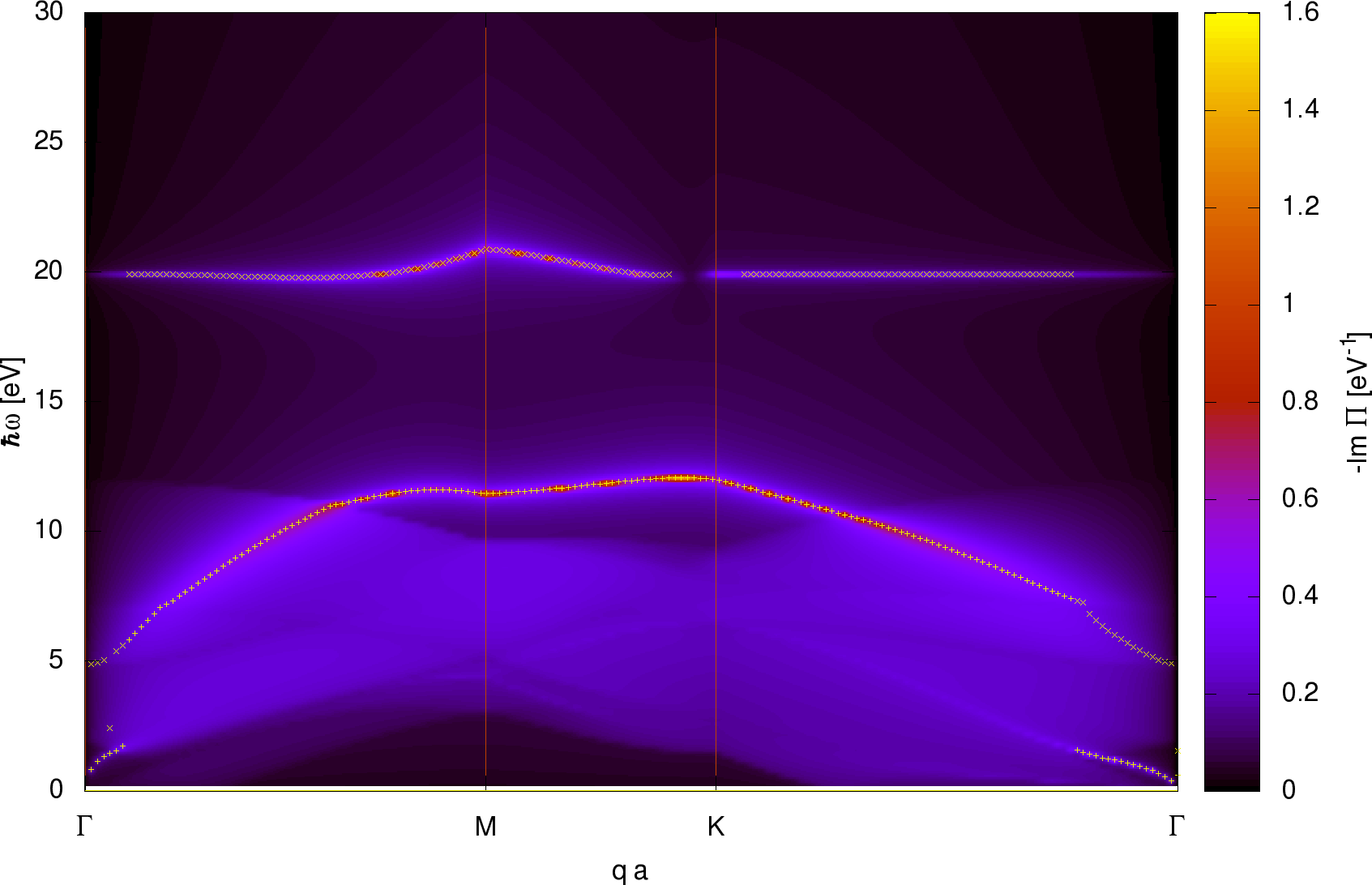}
\end{center}
\caption{(Color online) Plasmon dispersion relation for doped graphene at finite
temperature ($\mu=1$~eV, $T=3$~K), including LFE, with strain applied along
the $\theta=\pi/4$ (generic) direction. Strain increases (from left to right,
from top to bottom) as $\varepsilon = 0$, $0.075$, $0.175$, $0.275$.}
\label{fig:strain45}
\end{figure}

A qualitatively similar analysis applies to the case of strain applied along the
zig-zag direction ($\theta=\pi/6$, Fig.~\ref{fig:strain30}), and for strain
applied along a generic direction ($\theta=\pi/4$, Fig.~\ref{fig:strain45}),
with $\omega_2 (\bq)$ dispersing more weakly as the strain increases.

\begin{figure}[t]
\begin{center}
\includegraphics[width=0.45\columnwidth]{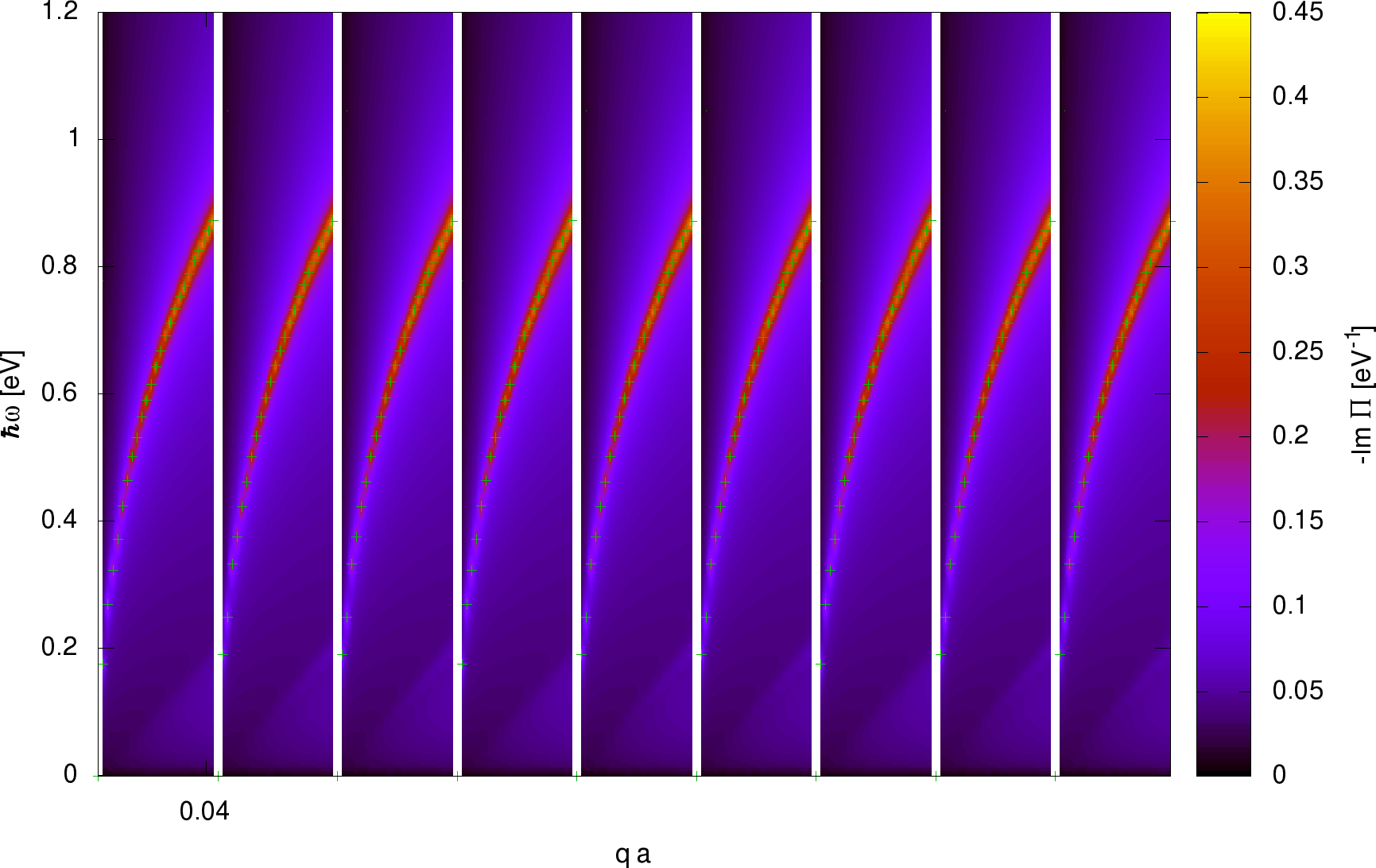}
\includegraphics[width=0.45\columnwidth]{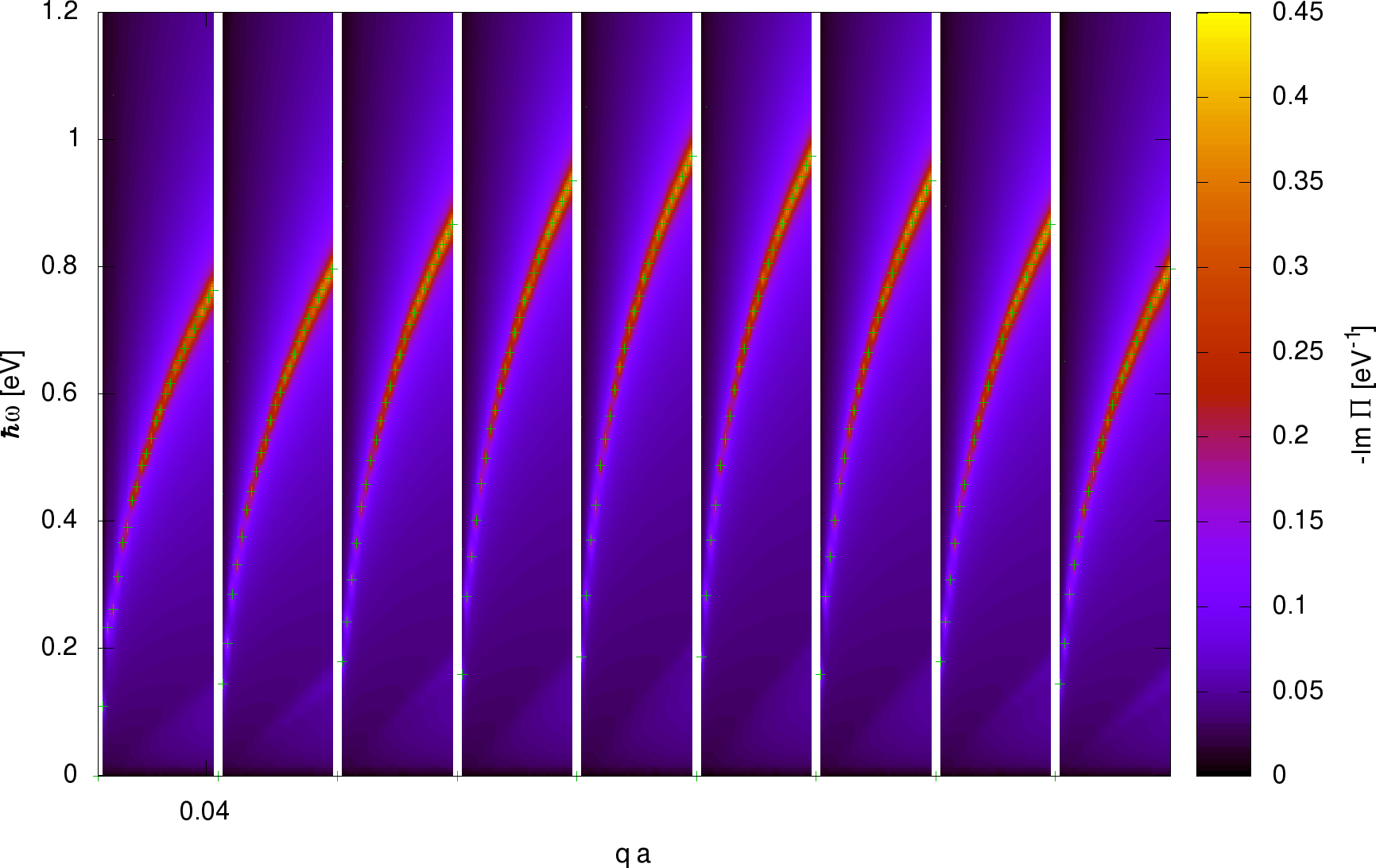}\\
\includegraphics[width=0.45\columnwidth]{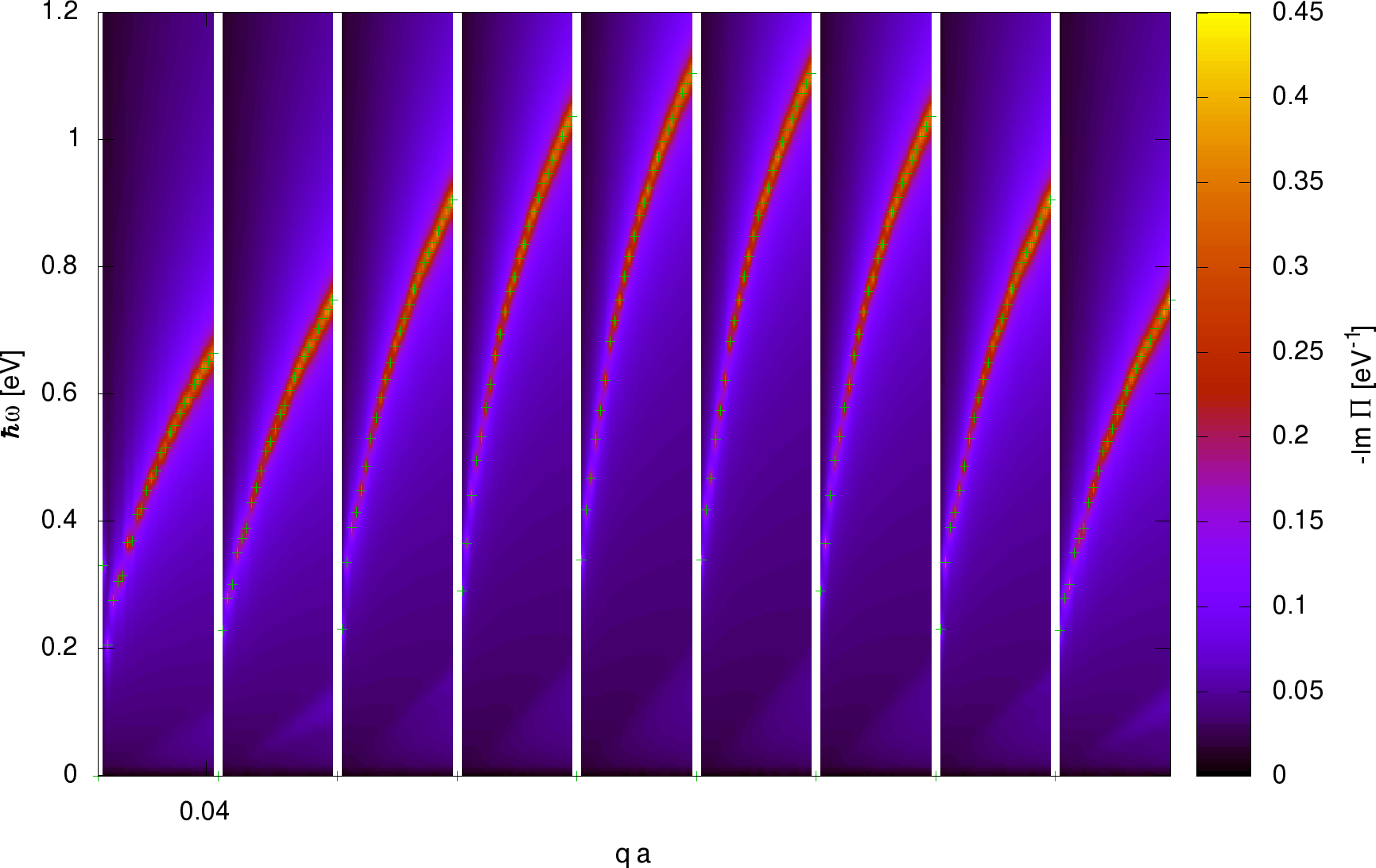}
\includegraphics[width=0.45\columnwidth]{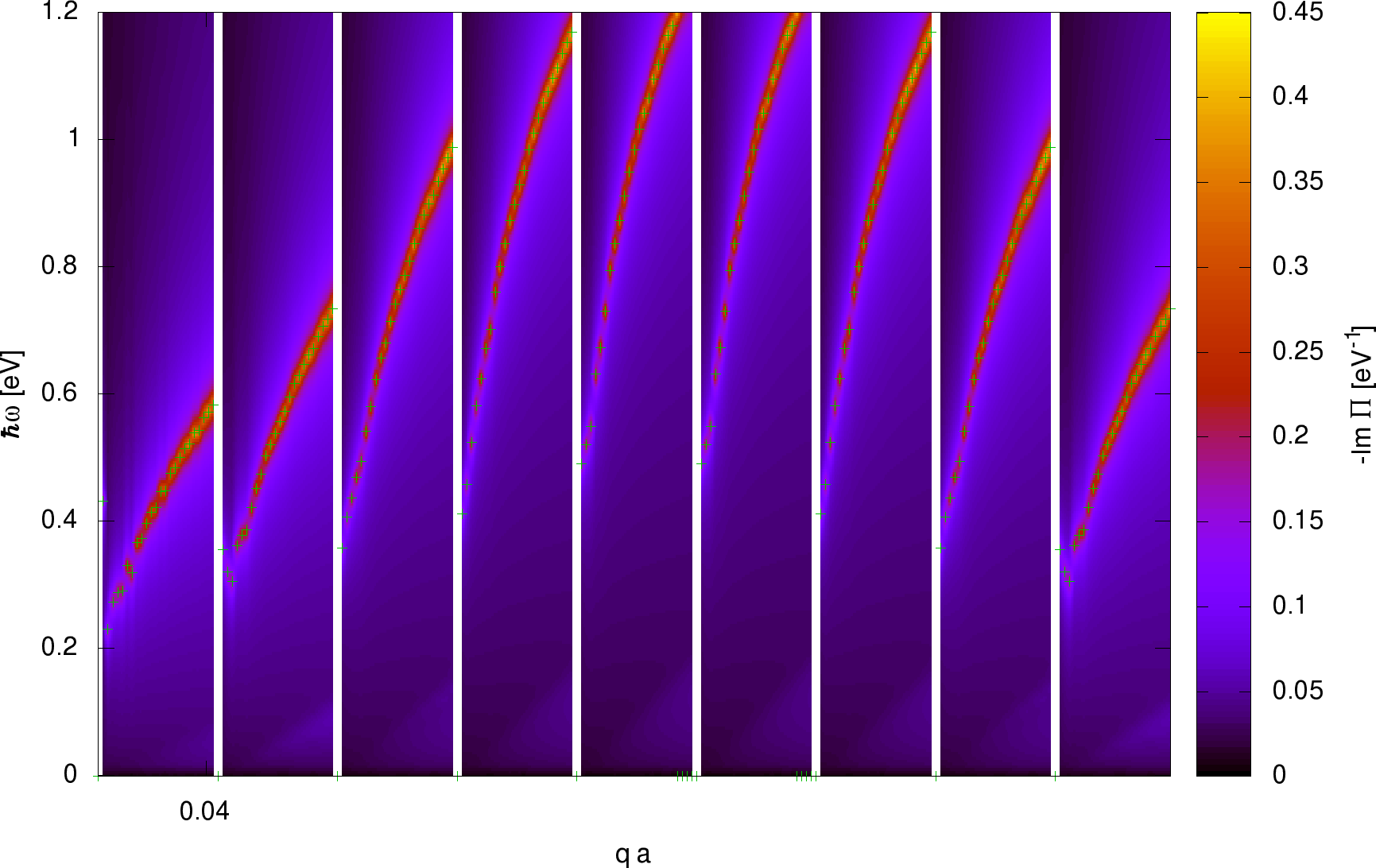}
\end{center}
\caption{(Color online) Plasmon dispersion relation for doped graphene at finite
temperature ($\mu=1$~eV, $T=3$~K), including LFE, with strain applied along
the $\theta=0$ (armchair) direction. Strain increases (from left to right, from
top to bottom) as $\varepsilon = 0$, $0.075$, $0.175$, $0.275$. In each graph,
different panels refer to $\omega_1 (\bq) = \omega_1(q,\varphi_\bq )$, with
$\varphi_\bq = 0^\circ, 20^\circ, 40^\circ , \ldots 160^\circ$.}
\label{fig:strainlowenergy0}
\end{figure}

Finally, we turn to study the $\bq$-dependence of the low-frequency, `acoustic'
mode $\omega_1 (\bq) \equiv \omega_1 (q,\varphi_\bq )$ under applied strain,
where $q=|\bq|$ and $\varphi_\bq$ denotes the angle between $\bq$ and the $x$
axis. Fig.~\ref{fig:strainlowenergy0} shows then the dispersion relation of the
lower plasmon branch as a function of $q$ for several values of $\varphi_\bq$,
for increasing strain applied along the armchair direction ($\theta=0$). While
the overall square-root shape $\omega_1 \approx \tilde{\omega}_1 \sqrt{qa}$,
Eq.~(\ref{eq:omegaasymp}), is maintained in all cases, one observes a stiffening
of such plasmonic mode with increasing strain and a maximum of the coefficient
$\tilde{\omega}_1$, Eq.~(\ref{eq:alpha1}), when $\varphi_\bq -\theta \approx 90^\circ$.
The same description qualitatively applies also to the cases of strain applied
along the armchair ($\theta=\pi/6$), and along a generic ($\theta=\pi/4$)
direction. Such a behavior can be justified analytically in the limit of no LFE
(cf. Sec.~\ref{sec:asymptotics}), and corresponds to the strain dependence
obtained for the optical conductivity \cite{Pellegrino:09b}. Indeed, from
Eq.~(\ref{eq:alpha1}), one may notice that all the strain dependence is
contained in the modulus square of the quasiparticle dispersion relation of the
conduction band at the Fermi level, $|\nabla_\bq E_{\bk2} /a|$. One finds
\begin{equation}
\tilde{\omega}_1 \propto
\left| \nabla_\bq E_{\bq2} \right| = \left( \frac{\cos^2
(\varphi_\bq -\eta)}{A^2} + \frac{\sin^2 (\varphi_\bq -\eta)}{B^2}
\right)^{1/2} ,
\label{eq:thetaphi}
\end{equation}
where $A$ and $B$ denote the semiaxes of the constant energy ellipse
\cite{Pellegrino:09b},
\begin{subequations}
\begin{eqnarray}
A^{-2} &=& \frac{1}{2} (\gamma - \sqrt{\alpha^2 + \beta^2} ),\\
B^{-2} &=& \frac{1}{2} (\gamma + \sqrt{\alpha^2 + \beta^2} ),
\end{eqnarray}
\end{subequations}
with
\begin{subequations}
\begin{eqnarray}
\alpha &=& -\frac{3}{2} a^2 (t_1^2 + t_2^2 - 2 t_3^2 ),\\
\beta  &=& \frac{3\sqrt{3}}{2} a^2 (t_1^2 - t_2^2 ),\\
\gamma &=& \frac{3}{2} a^2 (t_1^2 + t_2^2 + t_3^2 ),
\end{eqnarray}
\end{subequations}
and
\begin{subequations}
\begin{eqnarray}
\cos(2\eta) &=& \frac{|\alpha|}{\sqrt{\alpha^2 + \beta^2}} ,\\
\sin(2\eta) &=& \frac{|\alpha|}{\alpha} \frac{\beta}{\sqrt{\alpha^2 + \beta^2}}.
\end{eqnarray}
\end{subequations}
It follows that $\tilde{\omega}_1$ attains its maximum values whenever
$\varphi_\bq -\eta = \pi/2$ (modulo $\pi$), and its minimum values whenever
$\varphi_\bq -\eta = 0$ (modulo $\pi$). It turns out that $\eta=\theta$ in the
zig-zag and armchair cases (cf. Fig.~\ref{fig:strainlowenergy0}), whereas
$\eta\simeq\theta$ in the generic case.

\section{Conclusions}
\label{sec:conclusions}

By studying the electronic polarization, we have derived the dispersion relation
of the plasmon modes in graphene. Besides including electron-electron
correlation at the RPA level, we have also considered local field effects, that
are specific to the peculiar lattice structure under study. As a consequence of
the two-band character of the electronic band structure of graphene, we find in
general two plasmonic branches: (1) a low-energy, acoustic-like one
with a square-root behavior at small wavevectors, and (2) a high-energy,
optical-like mode, weakly dispersing at small wavevectors. This is generic to
two-band systems, and might apply to other two-band systems as well, such as
MgB$_2$, and is analogous to the collective modes in two-dimensional quantum
wells. We also find an intermediate energy pseudo-plasmon mode, associated with
a logarithmic (\emph{viz.} non power-law) divergence of the polarization, which can
be related to an interband transition between the Van Hove singularities in the
valence and conduction bands of graphene, and can be identified with a
$\pi\to\pi^\ast$ transition. We have next studied, both analytically and
numerically, the dependence of the plasmon branches on applied strain. While the
square-root character of the low-energy mode at small wavevector is robust with
respect to applied strain, we find a nonmonotonic stiffening as a function of
the wavevector direction, the maximum steepness occurring roughly when
the latter is orthogonal to the direction of applied strain. We have also
studied the influence of applied strain on the high-energy, optical-like plasmon
branch.

\begin{small} \bibliographystyle{apsrev}
\bibliography{a,b,c,d,e,f,g,h,i,j,k,l,m,n,o,p,q,r,s,t,u,v,w,x,y,z,zzproceedings,Angilella,notes}

\begin{thebibliography}{45}
\expandafter\ifx\csname natexlab\endcsname\relax\def\natexlab#1{#1}\fi
\expandafter\ifx\csname bibnamefont\endcsname\relax
  \def\bibnamefont#1{#1}\fi
\expandafter\ifx\csname bibfnamefont\endcsname\relax
  \def\bibfnamefont#1{#1}\fi
\expandafter\ifx\csname citenamefont\endcsname\relax
  \def\citenamefont#1{#1}\fi
\expandafter\ifx\csname url\endcsname\relax
  \def\url#1{\texttt{#1}}\fi
\expandafter\ifx\csname urlprefix\endcsname\relax\def\urlprefix{URL }\fi
\providecommand{\bibinfo}[2]{#2}
\providecommand{\eprint}[2][]{\url{#2}}

\bibitem[{\citenamefont{Novoselov et~al.}(2005)\citenamefont{Novoselov, Jiang,
  Schedin, Booth, Khotkevich, Morozov, and Geim}}]{Novoselov:05a}
\bibinfo{author}{\bibfnamefont{K.~S.} \bibnamefont{Novoselov}},
  \bibinfo{author}{\bibfnamefont{D.}~\bibnamefont{Jiang}},
  \bibinfo{author}{\bibfnamefont{F.}~\bibnamefont{Schedin}},
  \bibinfo{author}{\bibfnamefont{T.~J.} \bibnamefont{Booth}},
  \bibinfo{author}{\bibfnamefont{V.~V.} \bibnamefont{Khotkevich}},
  \bibinfo{author}{\bibfnamefont{S.~V.} \bibnamefont{Morozov}},
  \bibnamefont{and} \bibinfo{author}{\bibfnamefont{A.~K.} \bibnamefont{Geim}},
  \bibinfo{journal}{Proc. Nat. Acad. Sci.} \textbf{\bibinfo{volume}{102}},
  \bibinfo{pages}{10451} (\bibinfo{year}{2005}).

\bibitem[{\citenamefont{{Castro Neto} et~al.}(2009)\citenamefont{{Castro Neto},
  Guinea, Peres, Novoselov, and Geim}}]{CastroNeto:08}
\bibinfo{author}{\bibfnamefont{A.~H.} \bibnamefont{{Castro Neto}}},
  \bibinfo{author}{\bibfnamefont{F.}~\bibnamefont{Guinea}},
  \bibinfo{author}{\bibfnamefont{N.~M.~R.} \bibnamefont{Peres}},
  \bibinfo{author}{\bibfnamefont{K.~S.} \bibnamefont{Novoselov}},
  \bibnamefont{and} \bibinfo{author}{\bibfnamefont{A.~K.} \bibnamefont{Geim}},
  \bibinfo{journal}{Rev. Mod. Phys.} \textbf{\bibinfo{volume}{81}},
  \bibinfo{pages}{000109} (\bibinfo{year}{2009}).

\bibitem[{\citenamefont{Abergel et~al.}(2010)\citenamefont{Abergel, Apalkov,
  Berashevich, Ziegler, and Chakraborty}}]{Abergel:10}
\bibinfo{author}{\bibfnamefont{D.~S.~L.} \bibnamefont{Abergel}},
  \bibinfo{author}{\bibfnamefont{V.}~\bibnamefont{Apalkov}},
  \bibinfo{author}{\bibfnamefont{J.}~\bibnamefont{Berashevich}},
  \bibinfo{author}{\bibfnamefont{K.}~\bibnamefont{Ziegler}}, \bibnamefont{and}
  \bibinfo{author}{\bibfnamefont{T.}~\bibnamefont{Chakraborty}},
  \bibinfo{journal}{Adv. Phys.} \textbf{\bibinfo{volume}{59}},
  \bibinfo{pages}{261} (\bibinfo{year}{2010}).

\bibitem[{\citenamefont{Zhang et~al.}(2005)\citenamefont{Zhang, Tan, Stormer,
  and Kim}}]{Zhang:05}
\bibinfo{author}{\bibfnamefont{Y.}~\bibnamefont{Zhang}},
  \bibinfo{author}{\bibfnamefont{Y.}~\bibnamefont{Tan}},
  \bibinfo{author}{\bibfnamefont{H.~L.} \bibnamefont{Stormer}},
  \bibnamefont{and} \bibinfo{author}{\bibfnamefont{P.}~\bibnamefont{Kim}},
  \bibinfo{journal}{Nature} \textbf{\bibinfo{volume}{438}},
  \bibinfo{pages}{201} (\bibinfo{year}{2005}).

\bibitem[{\citenamefont{Berger et~al.}(2006)\citenamefont{Berger, Song, Li, Wu,
  Brown, Naud, Mayou, Li, Hass, Marchenkov et~al.}}]{Berger:06}
\bibinfo{author}{\bibfnamefont{C.}~\bibnamefont{Berger}},
  \bibinfo{author}{\bibfnamefont{Z.}~\bibnamefont{Song}},
  \bibinfo{author}{\bibfnamefont{X.}~\bibnamefont{Li}},
  \bibinfo{author}{\bibfnamefont{X.}~\bibnamefont{Wu}},
  \bibinfo{author}{\bibfnamefont{N.}~\bibnamefont{Brown}},
  \bibinfo{author}{\bibfnamefont{C.}~\bibnamefont{Naud}},
  \bibinfo{author}{\bibfnamefont{D.}~\bibnamefont{Mayou}},
  \bibinfo{author}{\bibfnamefont{T.}~\bibnamefont{Li}},
  \bibinfo{author}{\bibfnamefont{J.}~\bibnamefont{Hass}},
  \bibinfo{author}{\bibfnamefont{A.~N.} \bibnamefont{Marchenkov}},
  \bibnamefont{et~al.}, \bibinfo{journal}{Science}
  \textbf{\bibinfo{volume}{312}}, \bibinfo{pages}{1191} (\bibinfo{year}{2006}).

\bibitem[{\citenamefont{Gonz\'alez et~al.}(1999)\citenamefont{Gonz\'alez,
  Guinea, and Vozmediano}}]{Gonzalez:99}
\bibinfo{author}{\bibfnamefont{J.}~\bibnamefont{Gonz\'alez}},
  \bibinfo{author}{\bibfnamefont{F.}~\bibnamefont{Guinea}}, \bibnamefont{and}
  \bibinfo{author}{\bibfnamefont{M.~A.~H.} \bibnamefont{Vozmediano}},
  \bibinfo{journal}{Phys. Rev. B} \textbf{\bibinfo{volume}{59}},
  \bibinfo{pages}{R2474} (\bibinfo{year}{1999}).

\bibitem[{\citenamefont{Vafek}(2006)}]{Vafek:06}
\bibinfo{author}{\bibfnamefont{O.}~\bibnamefont{Vafek}},
  \bibinfo{journal}{Phys. Rev. Lett.} \textbf{\bibinfo{volume}{97}},
  \bibinfo{pages}{266406} (\bibinfo{year}{2006}).

\bibitem[{\citenamefont{Wunsch et~al.}(2006)\citenamefont{Wunsch, Stauber,
  Sols, and Guinea}}]{Wunsch:06}
\bibinfo{author}{\bibfnamefont{B.}~\bibnamefont{Wunsch}},
  \bibinfo{author}{\bibfnamefont{T.}~\bibnamefont{Stauber}},
  \bibinfo{author}{\bibfnamefont{F.}~\bibnamefont{Sols}}, \bibnamefont{and}
  \bibinfo{author}{\bibfnamefont{F.}~\bibnamefont{Guinea}},
  \bibinfo{journal}{New Journal of Physics} \textbf{\bibinfo{volume}{8}},
  \bibinfo{pages}{318} (\bibinfo{year}{2006}),
  \urlprefix\url{http://stacks.iop.org/1367-2630/8/i=12/a=318}.

\bibitem[{\citenamefont{Hwang and Das~Sarma}(2007)}]{Hwang:07a}
\bibinfo{author}{\bibfnamefont{E.~H.} \bibnamefont{Hwang}} \bibnamefont{and}
  \bibinfo{author}{\bibfnamefont{S.}~\bibnamefont{Das~Sarma}},
  \bibinfo{journal}{Phys. Rev. B} \textbf{\bibinfo{volume}{75}},
  \bibinfo{pages}{205418} (\bibinfo{year}{2007}).

\bibitem[{\citenamefont{Stauber et~al.}(2010)\citenamefont{Stauber, Schliemann,
  and Peres}}]{Stauber:10}
\bibinfo{author}{\bibfnamefont{T.}~\bibnamefont{Stauber}},
  \bibinfo{author}{\bibfnamefont{J.}~\bibnamefont{Schliemann}},
  \bibnamefont{and} \bibinfo{author}{\bibfnamefont{N.~M.~R.}
  \bibnamefont{Peres}}, \bibinfo{journal}{Phys. Rev. B}
  \textbf{\bibinfo{volume}{81}}, \bibinfo{pages}{085409}
  (\bibinfo{year}{2010}).

\bibitem[{\citenamefont{Wang and Chakraborty}(2007)}]{Wang:07}
\bibinfo{author}{\bibfnamefont{X.-F.} \bibnamefont{Wang}} \bibnamefont{and}
  \bibinfo{author}{\bibfnamefont{T.}~\bibnamefont{Chakraborty}},
  \bibinfo{journal}{Phys. Rev. B} \textbf{\bibinfo{volume}{75}},
  \bibinfo{pages}{033408} (\bibinfo{year}{2007}).

\bibitem[{\citenamefont{Adler}(1962)}]{Adler:62}
\bibinfo{author}{\bibfnamefont{S.~L.} \bibnamefont{Adler}},
  \bibinfo{journal}{Phys. Rev.} \textbf{\bibinfo{volume}{126}},
  \bibinfo{pages}{413} (\bibinfo{year}{1962}).

\bibitem[{\citenamefont{{van Schilfgaarde} and
  Katsnelson}(2010)}]{Schilfgaarde:10}
\bibinfo{author}{\bibfnamefont{M.}~\bibnamefont{{van Schilfgaarde}}}
  \bibnamefont{and} \bibinfo{author}{\bibfnamefont{M.~I.}
  \bibnamefont{Katsnelson}}, \bibinfo{journal}{...}
  \textbf{\bibinfo{volume}{..}}, \bibinfo{pages}{...} (\bibinfo{year}{2010}),
  \bibinfo{note}{preprint {\tt arXiv:1006.2426v1}}.

\bibitem[{\citenamefont{Bostwick et~al.}(2007)\citenamefont{Bostwick, Ohta,
  Seyller, Horn, and Rotenberg}}]{Bostwick:07}
\bibinfo{author}{\bibfnamefont{A.}~\bibnamefont{Bostwick}},
  \bibinfo{author}{\bibfnamefont{T.}~\bibnamefont{Ohta}},
  \bibinfo{author}{\bibfnamefont{T.}~\bibnamefont{Seyller}},
  \bibinfo{author}{\bibfnamefont{K.}~\bibnamefont{Horn}}, \bibnamefont{and}
  \bibinfo{author}{\bibfnamefont{E.}~\bibnamefont{Rotenberg}},
  \bibinfo{journal}{Nat. Phys.} \textbf{\bibinfo{volume}{3}},
  \bibinfo{pages}{36} (\bibinfo{year}{2007}).

\bibitem[{\citenamefont{Brar et~al.}(2010)\citenamefont{Brar, Wickenburg,
  Panlasigui, Park, Wehling, Zhang, Decker, Girit, Balatsky, Louie
  et~al.}}]{Brar:10}
\bibinfo{author}{\bibfnamefont{V.~W.} \bibnamefont{Brar}},
  \bibinfo{author}{\bibfnamefont{S.}~\bibnamefont{Wickenburg}},
  \bibinfo{author}{\bibfnamefont{M.}~\bibnamefont{Panlasigui}},
  \bibinfo{author}{\bibfnamefont{C.-H.} \bibnamefont{Park}},
  \bibinfo{author}{\bibfnamefont{T.~O.} \bibnamefont{Wehling}},
  \bibinfo{author}{\bibfnamefont{Y.}~\bibnamefont{Zhang}},
  \bibinfo{author}{\bibfnamefont{R.}~\bibnamefont{Decker}},
  \bibinfo{author}{\bibfnamefont{{\c{C}}.}~\bibnamefont{Girit}},
  \bibinfo{author}{\bibfnamefont{A.~V.} \bibnamefont{Balatsky}},
  \bibinfo{author}{\bibfnamefont{S.~G.} \bibnamefont{Louie}},
  \bibnamefont{et~al.}, \bibinfo{journal}{Phys. Rev. Lett.}
  \textbf{\bibinfo{volume}{104}}, \bibinfo{pages}{036805}
  (\bibinfo{year}{2010}).

\bibitem[{\citenamefont{Jablan et~al.}(2009)\citenamefont{Jablan, Buljan, and
  Solja\ifmmode \check{c}\else \v{c}\fi{}i\ifmmode~\acute{c}\else
  \'{c}\fi{}}}]{Jablan:09}
\bibinfo{author}{\bibfnamefont{M.}~\bibnamefont{Jablan}},
  \bibinfo{author}{\bibfnamefont{H.}~\bibnamefont{Buljan}}, \bibnamefont{and}
  \bibinfo{author}{\bibfnamefont{M.}~\bibnamefont{Solja\ifmmode \check{c}\else
  \v{c}\fi{}i\ifmmode~\acute{c}\else \'{c}\fi{}}}, \bibinfo{journal}{Phys. Rev.
  B} \textbf{\bibinfo{volume}{80}}, \bibinfo{pages}{245435}
  (\bibinfo{year}{2009}).

\bibitem[{\citenamefont{Pereira and Castro~Neto}(2009)}]{Pereira:09}
\bibinfo{author}{\bibfnamefont{V.~M.} \bibnamefont{Pereira}} \bibnamefont{and}
  \bibinfo{author}{\bibfnamefont{A.~H.} \bibnamefont{Castro~Neto}},
  \bibinfo{journal}{Phys. Rev. Lett.} \textbf{\bibinfo{volume}{103}},
  \bibinfo{pages}{046801} (\bibinfo{year}{2009}).

\bibitem[{\citenamefont{Booth et~al.}(2008)\citenamefont{Booth, Blake, Nair,
  Jiang, Hill, Bangert, Bleloch, Gass, Novoselov, Katsnelson
  et~al.}}]{Booth:08}
\bibinfo{author}{\bibfnamefont{T.~J.} \bibnamefont{Booth}},
  \bibinfo{author}{\bibfnamefont{P.}~\bibnamefont{Blake}},
  \bibinfo{author}{\bibfnamefont{R.~R.} \bibnamefont{Nair}},
  \bibinfo{author}{\bibfnamefont{D.}~\bibnamefont{Jiang}},
  \bibinfo{author}{\bibfnamefont{E.~W.} \bibnamefont{Hill}},
  \bibinfo{author}{\bibfnamefont{U.}~\bibnamefont{Bangert}},
  \bibinfo{author}{\bibfnamefont{A.}~\bibnamefont{Bleloch}},
  \bibinfo{author}{\bibfnamefont{M.}~\bibnamefont{Gass}},
  \bibinfo{author}{\bibfnamefont{K.~S.} \bibnamefont{Novoselov}},
  \bibinfo{author}{\bibfnamefont{M.~I.} \bibnamefont{Katsnelson}},
  \bibnamefont{et~al.}, \bibinfo{journal}{Nano Letters}
  \textbf{\bibinfo{volume}{8}}, \bibinfo{pages}{2442} (\bibinfo{year}{2008}).

\bibitem[{\citenamefont{Jiang et~al.}(2010)\citenamefont{Jiang, Wang, and
  Li}}]{Jiang:10}
\bibinfo{author}{\bibfnamefont{J.-W.} \bibnamefont{Jiang}},
  \bibinfo{author}{\bibfnamefont{J.-S.} \bibnamefont{Wang}}, \bibnamefont{and}
  \bibinfo{author}{\bibfnamefont{B.}~\bibnamefont{Li}}, \bibinfo{journal}{Phys.
  Rev. B} \textbf{\bibinfo{volume}{81}}, \bibinfo{pages}{073405}
  (\bibinfo{year}{2010}).

\bibitem[{\citenamefont{Choi et~al.}(2010)\citenamefont{Choi, Jhi, and
  Son}}]{Choi:10}
\bibinfo{author}{\bibfnamefont{S.-M.} \bibnamefont{Choi}},
  \bibinfo{author}{\bibfnamefont{S.-H.} \bibnamefont{Jhi}}, \bibnamefont{and}
  \bibinfo{author}{\bibfnamefont{Y.-W.} \bibnamefont{Son}},
  \bibinfo{journal}{Phys. Rev. B} \textbf{\bibinfo{volume}{81}},
  \bibinfo{pages}{081407} (\bibinfo{year}{2010}).

\bibitem[{\citenamefont{Cadelano et~al.}(2009)\citenamefont{Cadelano, Palla,
  Giordano, and Colombo}}]{Cadelano:09}
\bibinfo{author}{\bibfnamefont{E.}~\bibnamefont{Cadelano}},
  \bibinfo{author}{\bibfnamefont{P.~L.} \bibnamefont{Palla}},
  \bibinfo{author}{\bibfnamefont{S.}~\bibnamefont{Giordano}}, \bibnamefont{and}
  \bibinfo{author}{\bibfnamefont{L.}~\bibnamefont{Colombo}},
  \bibinfo{journal}{Phys. Rev. Lett.} \textbf{\bibinfo{volume}{102}},
  \bibinfo{pages}{235502} (\bibinfo{year}{2009}).

\bibitem[{\citenamefont{Liu et~al.}(2007)\citenamefont{Liu, Ming, and
  Li}}]{Liu:07}
\bibinfo{author}{\bibfnamefont{F.}~\bibnamefont{Liu}},
  \bibinfo{author}{\bibfnamefont{P.}~\bibnamefont{Ming}}, \bibnamefont{and}
  \bibinfo{author}{\bibfnamefont{J.}~\bibnamefont{Li}}, \bibinfo{journal}{Phys.
  Rev. B} \textbf{\bibinfo{volume}{76}}, \bibinfo{pages}{064120}
  (\bibinfo{year}{2007}).

\bibitem[{\citenamefont{Kim et~al.}(2009)\citenamefont{Kim, Zhao, Jang, Lee,
  Kim, Kim, Ahn, Kim, Choi, and Hong}}]{Kim:09}
\bibinfo{author}{\bibfnamefont{K.~S.} \bibnamefont{Kim}},
  \bibinfo{author}{\bibfnamefont{Y.}~\bibnamefont{Zhao}},
  \bibinfo{author}{\bibfnamefont{H.}~\bibnamefont{Jang}},
  \bibinfo{author}{\bibfnamefont{S.~Y.} \bibnamefont{Lee}},
  \bibinfo{author}{\bibfnamefont{J.~M.} \bibnamefont{Kim}},
  \bibinfo{author}{\bibfnamefont{K.~S.} \bibnamefont{Kim}},
  \bibinfo{author}{\bibfnamefont{J.~H.} \bibnamefont{Ahn}},
  \bibinfo{author}{\bibfnamefont{P.}~\bibnamefont{Kim}},
  \bibinfo{author}{\bibfnamefont{J.}~\bibnamefont{Choi}}, \bibnamefont{and}
  \bibinfo{author}{\bibfnamefont{B.~H.} \bibnamefont{Hong}},
  \bibinfo{journal}{Nature} \textbf{\bibinfo{volume}{457}},
  \bibinfo{pages}{706} (\bibinfo{year}{2009}).

\bibitem[{\citenamefont{Cocco et~al.}(2010)\citenamefont{Cocco, Cadelano, and
  Colombo}}]{Cocco:10}
\bibinfo{author}{\bibfnamefont{G.}~\bibnamefont{Cocco}},
  \bibinfo{author}{\bibfnamefont{E.}~\bibnamefont{Cadelano}}, \bibnamefont{and}
  \bibinfo{author}{\bibfnamefont{L.}~\bibnamefont{Colombo}},
  \bibinfo{journal}{Phys. Rev. B} \textbf{\bibinfo{volume}{81}},
  \bibinfo{pages}{241412} (\bibinfo{year}{2010}).

\bibitem[{\citenamefont{Pereira et~al.}(2009)\citenamefont{Pereira, {Castro
  Neto}, and Peres}}]{Pereira:08a}
\bibinfo{author}{\bibfnamefont{V.~M.} \bibnamefont{Pereira}},
  \bibinfo{author}{\bibfnamefont{A.~H.} \bibnamefont{{Castro Neto}}},
  \bibnamefont{and} \bibinfo{author}{\bibfnamefont{N.~M.~R.}
  \bibnamefont{Peres}}, \bibinfo{journal}{Phys. Rev. B}
  \textbf{\bibinfo{volume}{80}}, \bibinfo{pages}{045401}
  (\bibinfo{year}{2009}), \bibinfo{note}{preprint {\tt arXiv:0811.4396}}.

\bibitem[{\citenamefont{Gui et~al.}(2008)\citenamefont{Gui, Li, and
  Zhong}}]{Gui:08}
\bibinfo{author}{\bibfnamefont{G.}~\bibnamefont{Gui}},
  \bibinfo{author}{\bibfnamefont{J.}~\bibnamefont{Li}}, \bibnamefont{and}
  \bibinfo{author}{\bibfnamefont{J.}~\bibnamefont{Zhong}},
  \bibinfo{journal}{Phys. Rev. B} \textbf{\bibinfo{volume}{78}},
  \bibinfo{pages}{075435} (\bibinfo{year}{2008}).

\bibitem[{\citenamefont{Ribeiro et~al.}(2009)\citenamefont{Ribeiro, Pereira,
  Peres, Briddon, and Neto}}]{Ribeiro:09}
\bibinfo{author}{\bibfnamefont{R.~M.} \bibnamefont{Ribeiro}},
  \bibinfo{author}{\bibfnamefont{V.~M.} \bibnamefont{Pereira}},
  \bibinfo{author}{\bibfnamefont{N.~M.~R.} \bibnamefont{Peres}},
  \bibinfo{author}{\bibfnamefont{P.~R.} \bibnamefont{Briddon}},
  \bibnamefont{and} \bibinfo{author}{\bibfnamefont{A.~H.~C.}
  \bibnamefont{Neto}}, \bibinfo{journal}{New J. Phys.}
  \textbf{\bibinfo{volume}{11}}, \bibinfo{pages}{115002}
  (\bibinfo{year}{2009}).

\bibitem[{\citenamefont{Pellegrino
  et~al.}(2010{\natexlab{a}})\citenamefont{Pellegrino, Angilella, and
  Pucci}}]{Pellegrino:09b}
\bibinfo{author}{\bibfnamefont{F.~M.~D.} \bibnamefont{Pellegrino}},
  \bibinfo{author}{\bibfnamefont{G.~G.~N.} \bibnamefont{Angilella}},
  \bibnamefont{and} \bibinfo{author}{\bibfnamefont{R.}~\bibnamefont{Pucci}},
  \bibinfo{journal}{Phys. Rev. B} \textbf{\bibinfo{volume}{81}},
  \bibinfo{pages}{035411} (\bibinfo{year}{2010}{\natexlab{a}}).

\bibitem[{\citenamefont{Pellegrino
  et~al.}(2009{\natexlab{a}})\citenamefont{Pellegrino, Angilella, and
  Pucci}}]{Pellegrino:09c}
\bibinfo{author}{\bibfnamefont{F.~M.~D.} \bibnamefont{Pellegrino}},
  \bibinfo{author}{\bibfnamefont{G.~G.~N.} \bibnamefont{Angilella}},
  \bibnamefont{and} \bibinfo{author}{\bibfnamefont{R.}~\bibnamefont{Pucci}},
  \bibinfo{journal}{High Press. Res.} \textbf{\bibinfo{volume}{29}},
  \bibinfo{pages}{569} (\bibinfo{year}{2009}{\natexlab{a}}).

\bibitem[{opt()}]{optical-lattices}
\bibinfo{note}{It is relevant to add that deformed honeycomb lattices may be
  realized not only in strained graphene, but also in suitably designed optical
  lattices \cite{Bahat-Treidel:10}. In that context, it has been emphasized
  that as a function of strain, there exists a critical deformation beyond
  which a gap opens in the energy spectrum of both material (graphene) and
  optical lattices, and that Klein tunneling loses its relativistic character
  beyond the critical deformation, depending on the direction of applied
  strain.}

\bibitem[{\citenamefont{Reich et~al.}(2002)\citenamefont{Reich, Maultzsch,
  Thomsen, and Ordej\'on}}]{Reich:02}
\bibinfo{author}{\bibfnamefont{S.}~\bibnamefont{Reich}},
  \bibinfo{author}{\bibfnamefont{J.}~\bibnamefont{Maultzsch}},
  \bibinfo{author}{\bibfnamefont{C.}~\bibnamefont{Thomsen}}, \bibnamefont{and}
  \bibinfo{author}{\bibfnamefont{P.}~\bibnamefont{Ordej\'on}},
  \bibinfo{journal}{Phys. Rev. B} \textbf{\bibinfo{volume}{66}},
  \bibinfo{pages}{035412} (\bibinfo{year}{2002}).

\bibitem[{\citenamefont{Pellegrino
  et~al.}(2009{\natexlab{b}})\citenamefont{Pellegrino, Angilella, and
  Pucci}}]{Pellegrino:09}
\bibinfo{author}{\bibfnamefont{F.~M.~D.} \bibnamefont{Pellegrino}},
  \bibinfo{author}{\bibfnamefont{G.~G.~N.} \bibnamefont{Angilella}},
  \bibnamefont{and} \bibinfo{author}{\bibfnamefont{R.}~\bibnamefont{Pucci}},
  \bibinfo{journal}{Phys. Rev. B} \textbf{\bibinfo{volume}{80}},
  \bibinfo{pages}{094203} (\bibinfo{year}{2009}{\natexlab{b}}).

\bibitem[{\citenamefont{Giuliani and Vignale}(2005)}]{Giuliani:05}
\bibinfo{author}{\bibfnamefont{G.}~\bibnamefont{Giuliani}} \bibnamefont{and}
  \bibinfo{author}{\bibfnamefont{G.}~\bibnamefont{Vignale}},
  \emph{\bibinfo{title}{Quantum Theory of the Electron Liquid}}
  (\bibinfo{publisher}{Cambridge University Press},
  \bibinfo{address}{Cambridge}, \bibinfo{year}{2005}).

\bibitem[{\citenamefont{Schattke}(2005)}]{Schattke:05}
\bibinfo{author}{\bibfnamefont{W.}~\bibnamefont{Schattke}}, in
  \emph{\bibinfo{booktitle}{Encyclopedia of Condensed Matter Physics}}, edited
  by \bibinfo{editor}{\bibfnamefont{F.}~\bibnamefont{Bassani}},
  \bibinfo{editor}{\bibfnamefont{G.~L.} \bibnamefont{Liedl}}, \bibnamefont{and}
  \bibinfo{editor}{\bibfnamefont{P.}~\bibnamefont{Wyder}}
  (\bibinfo{publisher}{Elsevier}, \bibinfo{address}{Amsterdam},
  \bibinfo{year}{2005}), vol.~\bibinfo{volume}{1}, p. \bibinfo{pages}{145}.

\bibitem[{\citenamefont{Hanke and Sham}(1974)}]{Hanke:74}
\bibinfo{author}{\bibfnamefont{W.}~\bibnamefont{Hanke}} \bibnamefont{and}
  \bibinfo{author}{\bibfnamefont{L.~J.} \bibnamefont{Sham}},
  \bibinfo{journal}{Phys. Rev. Lett.} \textbf{\bibinfo{volume}{33}},
  \bibinfo{pages}{582} (\bibinfo{year}{1974}).

\bibitem[{\citenamefont{Hanke and Sham}(1975)}]{Hanke:74a}
\bibinfo{author}{\bibfnamefont{W.}~\bibnamefont{Hanke}} \bibnamefont{and}
  \bibinfo{author}{\bibfnamefont{L.~J.} \bibnamefont{Sham}},
  \bibinfo{journal}{Phys. Rev. B} \textbf{\bibinfo{volume}{12}},
  \bibinfo{pages}{4501} (\bibinfo{year}{1975}).

\bibitem[{\citenamefont{Hill et~al.}(2009)\citenamefont{Hill, Mikhailov, and
  Ziegler}}]{Hill:09}
\bibinfo{author}{\bibfnamefont{A.}~\bibnamefont{Hill}},
  \bibinfo{author}{\bibfnamefont{S.~A.} \bibnamefont{Mikhailov}},
  \bibnamefont{and} \bibinfo{author}{\bibfnamefont{K.}~\bibnamefont{Ziegler}},
  \bibinfo{journal}{EPL (Europhysics Letters)} \textbf{\bibinfo{volume}{87}},
  \bibinfo{pages}{27005} (\bibinfo{year}{2009}),
  \urlprefix\url{http://stacks.iop.org/0295-5075/87/i=2/a=27005}.

\bibitem[{\citenamefont{Gass et~al.}(2008)\citenamefont{Gass, Bangert, Bleloch,
  Wang, Nair, and Geim}}]{Gass:08}
\bibinfo{author}{\bibfnamefont{M.~H.} \bibnamefont{Gass}},
  \bibinfo{author}{\bibfnamefont{U.}~\bibnamefont{Bangert}},
  \bibinfo{author}{\bibfnamefont{A.~L.} \bibnamefont{Bleloch}},
  \bibinfo{author}{\bibfnamefont{P.}~\bibnamefont{Wang}},
  \bibinfo{author}{\bibfnamefont{R.~R.} \bibnamefont{Nair}}, \bibnamefont{and}
  \bibinfo{author}{\bibfnamefont{A.~K.} \bibnamefont{Geim}},
  \bibinfo{journal}{Nature Nanotech.} \textbf{\bibinfo{volume}{3}},
  \bibinfo{pages}{676} (\bibinfo{year}{2008}).

\bibitem[{\citenamefont{Ullrich and Vignale}(2002)}]{Ullrich:02}
\bibinfo{author}{\bibfnamefont{C.~A.} \bibnamefont{Ullrich}} \bibnamefont{and}
  \bibinfo{author}{\bibfnamefont{G.}~\bibnamefont{Vignale}},
  \bibinfo{journal}{Phys. Rev. B} \textbf{\bibinfo{volume}{65}},
  \bibinfo{pages}{245102} (\bibinfo{year}{2002}).

\bibitem[{\citenamefont{Eberlein et~al.}(2008)\citenamefont{Eberlein, Bangert,
  Nair, Jones, Gass, Bleloch, Novoselov, Geim, and Briddon}}]{Eberlein:08}
\bibinfo{author}{\bibfnamefont{T.}~\bibnamefont{Eberlein}},
  \bibinfo{author}{\bibfnamefont{U.}~\bibnamefont{Bangert}},
  \bibinfo{author}{\bibfnamefont{R.~R.} \bibnamefont{Nair}},
  \bibinfo{author}{\bibfnamefont{R.}~\bibnamefont{Jones}},
  \bibinfo{author}{\bibfnamefont{M.}~\bibnamefont{Gass}},
  \bibinfo{author}{\bibfnamefont{A.~L.} \bibnamefont{Bleloch}},
  \bibinfo{author}{\bibfnamefont{K.~S.} \bibnamefont{Novoselov}},
  \bibinfo{author}{\bibfnamefont{A.}~\bibnamefont{Geim}}, \bibnamefont{and}
  \bibinfo{author}{\bibfnamefont{P.~R.} \bibnamefont{Briddon}},
  \bibinfo{journal}{Phys. Rev. B} \textbf{\bibinfo{volume}{77}},
  \bibinfo{pages}{233406} (\bibinfo{year}{2008}).

\bibitem[{\citenamefont{Polini et~al.}(2008)\citenamefont{Polini, Asgari,
  Borghi, Barlas, Pereg-Barnea, and MacDonald}}]{Polini:08}
\bibinfo{author}{\bibfnamefont{M.}~\bibnamefont{Polini}},
  \bibinfo{author}{\bibfnamefont{R.}~\bibnamefont{Asgari}},
  \bibinfo{author}{\bibfnamefont{G.}~\bibnamefont{Borghi}},
  \bibinfo{author}{\bibfnamefont{Y.}~\bibnamefont{Barlas}},
  \bibinfo{author}{\bibfnamefont{T.}~\bibnamefont{Pereg-Barnea}},
  \bibnamefont{and} \bibinfo{author}{\bibfnamefont{A.~H.}
  \bibnamefont{MacDonald}}, \bibinfo{journal}{Phys. Rev. B}
  \textbf{\bibinfo{volume}{77}}, \bibinfo{pages}{081411}
  (\bibinfo{year}{2008}).

\bibitem[{\citenamefont{Das~Sarma and Hwang}(2009)}]{DasSarma:09}
\bibinfo{author}{\bibfnamefont{S.}~\bibnamefont{Das~Sarma}} \bibnamefont{and}
  \bibinfo{author}{\bibfnamefont{E.~H.} \bibnamefont{Hwang}},
  \bibinfo{journal}{Phys. Rev. Lett.} \textbf{\bibinfo{volume}{102}},
  \bibinfo{pages}{206412} (\bibinfo{year}{2009}).

\bibitem[{\citenamefont{Polini et~al.}(2009)\citenamefont{Polini, {MacDonald},
  and Vignale}}]{Polini:09}
\bibinfo{author}{\bibfnamefont{M.}~\bibnamefont{Polini}},
  \bibinfo{author}{\bibfnamefont{A.~H.} \bibnamefont{{MacDonald}}},
  \bibnamefont{and} \bibinfo{author}{\bibfnamefont{G.}~\bibnamefont{Vignale}},
  \bibinfo{journal}{...} \textbf{\bibinfo{volume}{...}}, \bibinfo{pages}{...}
  (\bibinfo{year}{2009}), \bibinfo{note}{preprint {\tt arXiv:0901.4528v1}}.

\bibitem[{\citenamefont{Pellegrino
  et~al.}(2010{\natexlab{b}})\citenamefont{Pellegrino, Angilella, and
  Pucci}}]{Pellegrino:10c}
\bibinfo{author}{\bibfnamefont{F.~M.~D.} \bibnamefont{Pellegrino}},
  \bibinfo{author}{\bibfnamefont{G.~G.~N.} \bibnamefont{Angilella}},
  \bibnamefont{and} \bibinfo{author}{\bibfnamefont{R.}~\bibnamefont{Pucci}},
  \bibinfo{journal}{High Press. Res.} \textbf{\bibinfo{volume}{...}},
  \bibinfo{pages}{...} (\bibinfo{year}{2010}{\natexlab{b}}).

\bibitem[{\citenamefont{Bahat-Treidel et~al.}(2010)\citenamefont{Bahat-Treidel,
  Peleg, Grobman, Shapira, Segev, and Pereg-Barnea}}]{Bahat-Treidel:10}
\bibinfo{author}{\bibfnamefont{O.}~\bibnamefont{Bahat-Treidel}},
  \bibinfo{author}{\bibfnamefont{O.}~\bibnamefont{Peleg}},
  \bibinfo{author}{\bibfnamefont{M.}~\bibnamefont{Grobman}},
  \bibinfo{author}{\bibfnamefont{N.}~\bibnamefont{Shapira}},
  \bibinfo{author}{\bibfnamefont{M.}~\bibnamefont{Segev}}, \bibnamefont{and}
  \bibinfo{author}{\bibfnamefont{T.}~\bibnamefont{Pereg-Barnea}},
  \bibinfo{journal}{Phys. Rev. Lett.} \textbf{\bibinfo{volume}{104}},
  \bibinfo{pages}{063901} (\bibinfo{year}{2010}).

\end{thebibliography}
\end{small}

\end{document}